\documentclass[runningheads]{llncs}

\makeatletter
\@ifundefined{lhs2tex.lhs2tex.sty.read}%
  {\@namedef{lhs2tex.lhs2tex.sty.read}{}%
   \newcommand\SkipToFmtEnd{}%
   \newcommand\EndFmtInput{}%
   \long\def\SkipToFmtEnd#1\EndFmtInput{}%
  }\SkipToFmtEnd

\newcommand\ReadOnlyOnce[1]{\@ifundefined{#1}{\@namedef{#1}{}}\SkipToFmtEnd}
\usepackage{amstext}
\usepackage{amssymb}
\usepackage{stmaryrd}
\DeclareFontFamily{OT1}{cmtex}{}
\DeclareFontShape{OT1}{cmtex}{m}{n}
  {<5><6><7><8>cmtex8
   <9>cmtex9
   <10><10.95><12><14.4><17.28><20.74><24.88>cmtex10}{}
\DeclareFontShape{OT1}{cmtex}{m}{it}
  {<-> ssub * cmtt/m/it}{}

\DeclareFontShape{OT1}{cmtt}{bx}{n}
  {<5><6><7><8>cmtt8
   <9>cmbtt9
   <10><10.95><12><14.4><17.28><20.74><24.88>cmbtt10}{}
\DeclareFontShape{OT1}{cmtex}{bx}{n}
  {<-> ssub * cmtt/bx/n}{}

\newcommand{\Conid}[1]{\mathit{#1}}
\newcommand{\Varid}[1]{\mathit{#1}}
\newcommand{\anonymous}{\kern0.06em \vbox{\hrule\@width.5em}}

\renewcommand{\leq}{\leqslant}

\usepackage{polytable}

\@ifundefined{mathindent}%
  {\newdimen\mathindent\mathindent\leftmargini}%
  {}%

\def\resethooks{%
  \global\let\SaveRestoreHook\empty
  \global\let\ColumnHook\empty}
\newcommand*{\savecolumns}[1][default]%
  {\g@addto@macro\SaveRestoreHook{\savecolumns[#1]}}
\newcommand*{\restorecolumns}[1][default]%
  {\g@addto@macro\SaveRestoreHook{\restorecolumns[#1]}}
\newcommand*{\aligncolumn}[2]%
  {\g@addto@macro\ColumnHook{\column{#1}{#2}}}

\resethooks

\newcommand{\onelinecommentchars}{\quad-{}- }
\newcommand{\commentbeginchars}{\enskip\{-}
\newcommand{\commentendchars}{-\}\enskip}

\newcommand{\visiblecomments}{%
  \let\onelinecomment=\onelinecommentchars
  \let\commentbegin=\commentbeginchars
  \let\commentend=\commentendchars}

\newcommand{\invisiblecomments}{%
  \let\onelinecomment=\empty
  \let\commentbegin=\empty
  \let\commentend=\empty}

\visiblecomments

\newlength{\blanklineskip}
\newcommand{\hsindent}[1]{\quad}
\let\hspre\empty
\let\hspost\empty

\EndFmtInput
\makeatother
\ReadOnlyOnce{forall.fmt}%
\makeatletter

\let\HaskellResetHook\empty
\newcommand*{\AtHaskellReset}[1]{%
  \g@addto@macro\HaskellResetHook{#1}}
\newcommand*{\HaskellReset}{\HaskellResetHook}

\newcommand\hsforall{\global\let\hsdot=\hsperiodonce}
\newcommand*\hsperiodonce[2]{#2\global\let\hsdot=\hscompose}
\newcommand*\hscompose[2]{#1}

\AtHaskellReset{\global\let\hsdot=\hscompose}

\HaskellReset

\makeatother
\EndFmtInput
\ReadOnlyOnce{polycode.fmt}%
\makeatletter

\newcommand{\hsnewpar}[1]%
  {{\parskip=0pt\parindent=0pt\par\vskip #1\noindent}}

\newcommand{\hscodestyle}{}

\newcommand{\sethscode}[1]%
  {\expandafter\let\expandafter\hscode\csname #1\endcsname
   \expandafter\let\expandafter\endhscode\csname end#1\endcsname}

  {\par\noindent
   \advance\leftskip\mathindent
   \hscodestyle
   \let\\=\@normalcr
   \let\hspre\(\let\hspost\)%
   \pboxed}%
  {\endpboxed\)%
   \par\noindent
   \ignorespacesafterend}

  {\hsnewpar\abovedisplayskip
   \advance\leftskip\mathindent
   \hscodestyle
   \let\hspre\(\let\hspost\)%
   \pboxed}%
  {\endpboxed%
   \hsnewpar\belowdisplayskip
   \ignorespacesafterend}

  {\hsnewpar\abovedisplayskip
   \advance\leftskip\mathindent
   \hscodestyle
   \let\\=\@normalcr
   \(\pboxed}%
  {\endpboxed\)%
   \hsnewpar\belowdisplayskip
   \ignorespacesafterend}

\newcommand{\plainhs}{\sethscode{plainhscode}}

\plainhs

  {\hsnewpar\abovedisplayskip
   \advance\leftskip\mathindent
   \hscodestyle
   \let\\=\@normalcr
   \(\parray}%
  {\endparray\)%
   \hsnewpar\belowdisplayskip
   \ignorespacesafterend}

  {\parray}{\endparray}

  {\(\parray}{\endparray\)}

\def\codeframewidth{\arrayrulewidth}
\RequirePackage{calc}

  {\parskip=\abovedisplayskip\par\noindent
   \hscodestyle
   \arrayrulewidth=\codeframewidth
   \tabular{@{}|p{\linewidth-2\arraycolsep-2\arrayrulewidth-2pt}|@{}}%
   \hline\framedhslinecorrect\\{-1.5ex}%
   \let\endoflinesave=\\
   \let\\=\@normalcr
   \(\pboxed}%
  {\endpboxed\)%
   \framedhslinecorrect\endoflinesave{.5ex}\hline
   \endtabular
   \parskip=\belowdisplayskip\par\noindent
   \ignorespacesafterend}

\newcommand{\framedhslinecorrect}[2]%
  {#1[#2]}

  {\(\def\column##1##2{}%
   \let\>\undefined\let\<\undefined\let\\\undefined
   \newcommand\>[1][]{}\newcommand\<[1][]{}\newcommand\\[1][]{}%
   \def\fromto##1##2##3{##3}%
   }{\) }%

  {\let\orighscode=\hscode
   \let\origendhscode=\endhscode
   \def\endhscode{\def\hscode{\endgroup\def\@currenvir{hscode}\\}\begingroup}
   \orighscode\def\hscode{\endgroup\def\@currenvir{hscode}}}%
  {\origendhscode
   \global\let\hscode=\orighscode
   \global\let\endhscode=\origendhscode}%

\makeatother
\EndFmtInput

\ReadOnlyOnce{Formatting.fmt}%
\makeatletter

\let\Varid\mathit
\let\Conid\mathsf

\def\commentbegin{\quad\begingroup\color{Green}\{\ }
\def\commentend{\}\endgroup}

\makeatother
\EndFmtInput

\usepackage{amsmath}
\usepackage{mathpartir}
\usepackage{amsfonts}
\usepackage{stmaryrd}
\usepackage{hyperref}
\usepackage[dvipsnames]{xcolor}

\usepackage{scalerel}
\usepackage{bussproofs}
\EnableBpAbbreviations
\usepackage{url}
\usepackage{subfig}
\usepackage{enumitem}
\usepackage{mdframed}
\usepackage{multicol}
\usepackage{graphicx}
\usepackage{bm}

\usepackage{common/doubleequals}

\newcommand{\delete}[1]{}

\allowdisplaybreaks

\definecolor{mediumpersianblue}{rgb}{0.0, 0.4, 0.65}
\everymath{\color{mediumpersianblue}}

\begin{document}

\title{Declarative Pearl:
Deriving Monadic Quicksort%
\thanks{%
This is a post-peer-review, pre-copyedit version of an article published in Nakano K., Sagonas K. (eds) Functional and Logic Programming (FLOPS 2020), Springer.
The final authenticated version is available online at:
\url{https://doi.org/10.1007/978-3-030-59025-3_8}.
}
}
\author{Shin-Cheng Mu\inst{1} \and
Tsung-Ju Chiang\inst{2}}
%
%
\institute{Academia Sinica, Taiwan \and
 National Taiwan University, Taiwan}

%
\maketitle              
\begin{abstract}
To demonstrate derivation of monadic programs,
we present a specification of sorting using the non-determinism monad, and derive pure quicksort on lists and state-monadic quicksort on arrays.
In the derivation one may switch between point-free and pointwise styles, and deploy techniques familiar to functional programmers such as pattern matching and induction on structures or on sizes.
Derivation of stateful programs resembles reasoning backwards from the postcondition.
\end{abstract}

\section{Introduction}
\label{sec:intro}


This pearl presents two derivations of quicksort.
The purpose is to demonstrate reasoning and derivation of monadic programs.
In the first derivation we present a specification of sorting using the non-determinism monad, from which we derive a pure function that sorts a list.
In the second derivation we derive an imperative algorithm, expressed in terms of the state monad, that sorts an array.

Before we dive into the derivations, we shall explain our motivation.
Program derivation is the technique of formally constructing a program from a problem specification.
In functional derivation, the specification is a function that obviously matches the problem description, albeit inefficiently.
It is then stepwise transformed to a program that is efficient enough,
where every step is justified by mathematical properties guaranteeing that the program {\em equals} the specification, that is, for all inputs they compute exactly the same output.

It often happens, for certain problem, that several answers are equally preferred.
In sorting, for example, the array to be sorted might contain items with identical keys.
It would be inflexible, if not impossible, to decide in the specification how to resolve the tie: it is hard to predict how quicksort arranges items with identical keys before actually deriving quicksort.%
\footnote{Unless we confine ourselves to stable sorting.}
Such problems are better modelled as non-deterministic mappings from the input to all valid outputs. The derived program no longer equals but {\em refines} the specification.%
\footnote{This is standard in imperative program derivation ---
Dijkstra~\cite{Dijkstra:76:Discipline}
argued that we should take non-determinism as default and determinism as a special case.}

To cope with non-determinism, there was a trend in the 90's generalising from functions to relations~\cite{BackhousedeBruin:91:Relational,
BirddeMoor:97:Algebra}.
Although these relational calculi are, for advocates including the authors of this paper, concise and elegant, for those who were not following this line of development, these calculi are hard to comprehend and use.
People, in their first and often only exposure to the calculi, often complained that the notations are too bizarre, and reasoning with inequality (refinement) too complex.
One source of difficulties is that notations of relational calculus are usually \emph{point-free} --- that is, about composing relations instead of applying relations to arguments.
There have been attempts (e.g \cite{deMoorGibbons:00:Pointwise,BirdRabe:19:How}) designing {\em pointwise} notations, which functional programmers are more familiar with.
Proposals along this line tend to exhibit confusion
when functions are applied to non-deterministic values
 --- $\beta$-reduction and $\eta$-conversion do not hold.
One example~\cite{deMoorGibbons:00:Pointwise} is that
\ensuremath{(\lambda \Varid{x}\to \Varid{x}\mathbin{-}\Varid{x})\;(\mathrm{0}\mathbin{\talloblong}\mathrm{1})}, where \ensuremath{(\talloblong)} denotes non-deterministic choice, always yields \ensuremath{\mathrm{0}}, while \ensuremath{(\mathrm{0}\mathbin{\talloblong}\mathrm{1})\mathbin{-}(\mathrm{0}\mathbin{\talloblong}\mathrm{1})} could be \ensuremath{\mathrm{0}}, \ensuremath{\mathrm{1}}, or \ensuremath{\mathbin{-}\mathrm{1}}.

Preceding the development of relations for program derivation, another way to model non-determinism has gained popularity.
Monads~\cite{Moggi:89:Computational} were introduced into functional programming as a way to rigorously talk about side effects including IO, state, exception, and non-determinism.
Although they are considered one of the main obstacles in learning functional programming (in particular Haskell), monads have gained wide acceptance.
In this pearl we propose a calculus of program derivation based on monads --- essentially moving to a Kleisli category.
Problem specifications are given as Kleisli arrows for non-deterministic monads, to be refined to deterministic functional programs through calculation.
One of the benefits is that functional programmers may deploy techniques they are familiar with when reasoning about and deriving programs.
These include both point-free and pointwise reasoning, and induction on structures or sizes of data types.
An additional benefit of using monads is that we may talk about effects other than non-determinism. We demonstrate how to, from a specification of quicksort on lists, construct the imperative quicksort for arrays. All the derivations and theorems in this pearl are verified in the dependently typed programming language Agda.%
\footnote{\url{ https://scm.iis.sinica.edu.tw/home/2020/deriving-monadic-quicksort/}}

\section{Monads}
\label{sec:monads}

A monad consists of a type constructor \ensuremath{\Varid{m}\mathbin{::}\mathbin{*}\to \mathbin{*}} paired with two operators, can be modelled in Haskell as a type class:
\begin{hscode}\SaveRestoreHook
\column{B}{@{}>{\hspre}l<{\hspost}@{}}%
\column{5}{@{}>{\hspre}l<{\hspost}@{}}%
\column{13}{@{}>{\hspre}l<{\hspost}@{}}%
\column{19}{@{}>{\hspre}l<{\hspost}@{}}%
\column{22}{@{}>{\hspre}l<{\hspost}@{}}%
\column{E}{@{}>{\hspre}l<{\hspost}@{}}%
\>[B]{}\mathbf{class}\;\Conid{Monad}\;\Varid{m}\;\mathbf{where}{}\<[E]%
\\
\>[B]{}\hsindent{5}{}\<[5]%
\>[5]{}\{\cdot \}{}\<[13]%
\>[13]{}\mathbin{::}\Varid{a}\to {}\<[22]%
\>[22]{}\Varid{m}\;\Varid{a}{}\<[E]%
\\
\>[B]{}\hsindent{5}{}\<[5]%
\>[5]{}(\mathrel{\hstretch{0.7}{>\!\!>\!\!=}}){}\<[13]%
\>[13]{}\mathbin{::}\Varid{m}\;{}\<[19]%
\>[19]{}\Varid{a}\to (\Varid{a}\to \Varid{m}\;\Varid{b})\to \Varid{m}\;\Varid{b}~~.{}\<[E]%
\ColumnHook
\end{hscode}\resethooks
The operator \ensuremath{\{\cdot \}} is usually called $\Varid{return}$ or $\Varid{unit}$.
Since it is used pervasively in this pearl, we use a shorter notation for brevity.
One can either think of it as mimicking the notation for a singleton
set, or C-style syntax for a block of effectful program.
They should satisfy the following \emph{monad laws}:
\begin{align*}
  \ensuremath{\Varid{m}\mathrel{\hstretch{0.7}{>\!\!>\!\!=}}\{\cdot \}} &= \ensuremath{\Varid{m}} \mbox{~~,}\\
  \ensuremath{\{\Varid{x}\}\mathrel{\hstretch{0.7}{>\!\!>\!\!=}}\Varid{f}} &= \ensuremath{\Varid{f}\;\Varid{x}} \mbox{~~,} \\
  \ensuremath{(\Varid{m}\mathrel{\hstretch{0.7}{>\!\!>\!\!=}}\Varid{f})\mathrel{\hstretch{0.7}{>\!\!>\!\!=}}\Varid{g}} &= \ensuremath{\Varid{m}\mathrel{\hstretch{0.7}{>\!\!>\!\!=}}(\lambda \Varid{x}\to \Varid{f}\;\Varid{x}\mathrel{\hstretch{0.7}{>\!\!>\!\!=}}\Varid{g})} \mbox{~~.}
\end{align*}

A standard operator \ensuremath{(\mathbin{\hstretch{0.7}{>\!\!>}})\mathbin{::}\Conid{Monad}\;\Varid{m}\Rightarrow \Varid{m}\;\Varid{a}\to \Varid{m}\;\Varid{b}\to \Varid{m}\;\Varid{b}}, defined by \ensuremath{\Varid{m}_{1}\mathbin{\hstretch{0.7}{>\!\!>}}\Varid{m}_{2}\mathrel{=}\Varid{m}_{1}\mathrel{\hstretch{0.7}{>\!\!>\!\!=}}\lambda \anonymous \to \Varid{m}_{2}}, is handy when we do not need the result of \ensuremath{\Varid{m}_{1}}.
Monadic functions can be combined by Kleisli composition \ensuremath{(\mathrel{\hstretch{0.7}{>\!\!=\!\!\!>}})},
defined by \ensuremath{\Varid{f}\mathrel{\hstretch{0.7}{>\!\!=\!\!\!>}}\Varid{g}~\mathrel{=}~\lambda \Varid{x}\to \Varid{f}\;\Varid{x}\mathrel{\hstretch{0.7}{>\!\!>\!\!=}}\Varid{g}}.
%

Monads usually come with additional operators corresponding to the effects they provide.
Regarding non-determinism, we assume two operators \ensuremath{\emptyset} and \ensuremath{(\talloblong)}, respectively denoting failure and non-deterministic choice:
\begin{hscode}\SaveRestoreHook
\column{B}{@{}>{\hspre}l<{\hspost}@{}}%
\column{3}{@{}>{\hspre}l<{\hspost}@{}}%
\column{10}{@{}>{\hspre}l<{\hspost}@{}}%
\column{E}{@{}>{\hspre}l<{\hspost}@{}}%
\>[B]{}\mathbf{class}\;\Conid{Monad}\;\Varid{m}\Rightarrow \Conid{MonadPlus}\;\Varid{m}\;\mathbf{where}{}\<[E]%
\\
\>[B]{}\hsindent{3}{}\<[3]%
\>[3]{}\emptyset{}\<[10]%
\>[10]{}\mathbin{::}\Varid{m}\;\Varid{a}{}\<[E]%
\\
\>[B]{}\hsindent{3}{}\<[3]%
\>[3]{}(\talloblong){}\<[10]%
\>[10]{}\mathbin{::}\Varid{m}\;\Varid{a}\to \Varid{m}\;\Varid{a}\to \Varid{m}\;\Varid{a}~~.{}\<[E]%
\ColumnHook
\end{hscode}\resethooks
It might be a good time to note that this pearl uses type classes for two purposes: firstly, to be explicit about the effects a program uses.
Secondly, the notation implies that it does not matter which actual implementation we use for \ensuremath{\Varid{m}}, as long as it satisfies all the properties we demand ---
as Gibbons and Hinze \cite{GibbonsHinze:11:Just} proposed, we use the properties, not the implementations, when reasoning about programs.
The style of reasoning in this pearl is not tied to type classes or Haskell,
and we do not strictly follow the particularities of type classes in the current Haskell standard.%
\footnote{For example, we overlook that a \ensuremath{\Conid{Monad}} must also be \ensuremath{\Conid{Applicative}}, \ensuremath{\Conid{MonadPlus}} be \ensuremath{\Conid{Alternative}}, and that functional dependency is needed in a number of places.}

It is usually assumed that \ensuremath{(\talloblong)} is associative with \ensuremath{\emptyset} as its identity:
\begin{align*}
  \ensuremath{\emptyset\mathbin{\talloblong}\Varid{m}} & = \ensuremath{\Varid{m}} ~=~ \ensuremath{\Varid{m}\mathbin{\talloblong}\emptyset} \mbox{~~,} &
  \ensuremath{(\Varid{m}_{1}\mathbin{\talloblong}\Varid{m}_{2})\mathbin{\talloblong}\Varid{m}_{3}} &=
     \ensuremath{\Varid{m}_{1}\mathbin{\talloblong}(\Varid{m}_{2}\mathbin{\talloblong}\Varid{m}_{3})} \mbox{~~.}
\end{align*}
For the purpose of this pearl, we also demand that \ensuremath{(\talloblong)} be idempotent and commutative. That is, \ensuremath{\Varid{m}\mathbin{\talloblong}\Varid{m}\mathrel{=}\Varid{m}} and \ensuremath{\Varid{m}\mathbin{\talloblong}\Varid{n}\mathrel{=}\Varid{n}\mathbin{\talloblong}\Varid{m}}.
Efficient implementations of such monads have been proposed (e.g. \cite{Kiselyov:13:How}).
However, we use non-determinism monad only in specification.
The derived programs are always deterministic.

The laws below concern interaction between non-determinism and \ensuremath{(\mathrel{\hstretch{0.7}{>\!\!>\!\!=}})}:
\begin{align}
  \ensuremath{\emptyset\mathrel{\hstretch{0.7}{>\!\!>\!\!=}}\Varid{f}} & = \ensuremath{\emptyset} \label{eq:nd-left-zero}\mbox{~~,}\\
  \ensuremath{\Varid{m}\mathbin{\hstretch{0.7}{>\!\!>}}\emptyset} & = \ensuremath{\emptyset} \label{eq:nd-right-zero}\mbox{~~,}\\
  \ensuremath{(\Varid{m}_{1}\mathbin{\talloblong}\Varid{m}_{2})\mathrel{\hstretch{0.7}{>\!\!>\!\!=}}\Varid{f}} &= \ensuremath{(\Varid{m}_{1}\mathrel{\hstretch{0.7}{>\!\!>\!\!=}}\Varid{f})\mathbin{\talloblong}(\Varid{m}_{2}\mathrel{\hstretch{0.7}{>\!\!>\!\!=}}\Varid{f})} \mbox{~~,}
     \label{eq:nd-left-distr}\\
  \ensuremath{\Varid{m}\mathrel{\hstretch{0.7}{>\!\!>\!\!=}}(\lambda \Varid{x}\to \Varid{f}_{1}\;\Varid{x}\mathbin{\talloblong}\Varid{f}_{2}\;\Varid{x})} &= \ensuremath{(\Varid{m}\mathrel{\hstretch{0.7}{>\!\!>\!\!=}}\Varid{f}_{1})\mathbin{\talloblong}(\Varid{m}\mathrel{\hstretch{0.7}{>\!\!>\!\!=}}\Varid{f}_{2})} \mbox{~~.}
     \label{eq:nd-right-distr}
\end{align}
Left-zero~\eqref{eq:nd-left-zero} and left-distributivity~\eqref{eq:nd-left-distr} are standard --- the latter says that \ensuremath{(\talloblong)} is algebraic.
When mixed with state, right-zero \eqref{eq:nd-right-zero} and right-distributivity \eqref{eq:nd-right-distr} imply that each non-deterministic branch has its own copy of the state~\cite{Pauwels:19:Handling}.

\section{Specification}
\label{sec:spec}

We are now ready to present a monadic specification of sorting.
Bird~\cite{Bird:96:Functional} demonstrated how to derive various sorting algorithms from relational specifications.
In Section \ref{sec:quicksort} and \ref{sec:iqsort} we show how quicksort can be derived in our monadic calculus.

We assume a type \ensuremath{\Conid{Elm}} (for ``elements'') associated with a total preorder \ensuremath{(\leq )}.
To sort a list \ensuremath{\Varid{xs}\mathbin{::}\Conid{List}\;\Conid{Elm}} is to choose, among all permutation of \ensuremath{\Varid{xs}}, those that are sorted:
\begin{hscode}\SaveRestoreHook
\column{B}{@{}>{\hspre}l<{\hspost}@{}}%
\column{E}{@{}>{\hspre}l<{\hspost}@{}}%
\>[B]{}\Varid{slowsort}\mathbin{::}\Conid{MonadPlus}\;\Varid{m}\Rightarrow \Conid{List}\;\Conid{Elm}\to \Varid{m}\;(\Conid{List}\;\Conid{Elm}){}\<[E]%
\\
\>[B]{}\Varid{slowsort}\mathrel{=}\Varid{perm}\mathrel{\hstretch{0.7}{>\!\!=\!\!\!>}}\Varid{filt}\;\Varid{sorted}~~,{}\<[E]%
\ColumnHook
\end{hscode}\resethooks
where \ensuremath{\Varid{perm}\mathbin{::}\Conid{MonadPlus}\;\Varid{m}\Rightarrow \Conid{List}\;\Varid{a}\to \Varid{m}\;(\Conid{List}\;\Varid{a})} non-deterministically computes a permutation of its input, \ensuremath{\Varid{sorted}\mathbin{::}\Conid{List}\;\Conid{Elm}\to \Conid{Bool}} checks whether a list is sorted, and \ensuremath{\Varid{filt}\;\Varid{p}\;\Varid{x}} returns \ensuremath{\Varid{x}} if \ensuremath{\Varid{p}\;\Varid{x}} holds, and fails otherwise:
\begin{hscode}\SaveRestoreHook
\column{B}{@{}>{\hspre}l<{\hspost}@{}}%
\column{E}{@{}>{\hspre}l<{\hspost}@{}}%
\>[B]{}\Varid{filt}\mathbin{::}\Conid{MonadPlus}\;\Varid{m}\Rightarrow (\Varid{a}\to \Conid{Bool})\to \Varid{a}\to \Varid{m}\;\Varid{a}{}\<[E]%
\\
\>[B]{}\Varid{filt}\;\Varid{p}\;\Varid{x}\mathrel{=}\Varid{guard}\;(\Varid{p}\;\Varid{x})\mathbin{\hstretch{0.7}{>\!\!>}}\{\Varid{x}\}~~.{}\<[E]%
\ColumnHook
\end{hscode}\resethooks
The function \ensuremath{\Varid{guard}\;\Varid{b}\mathrel{=}\mathbf{if}\;\Varid{b}\;\mathbf{then}\;\{\}\;\mathbf{else}\;\emptyset} is standard.
%
The predicate \ensuremath{\Varid{sorted}\mathbin{::}\Conid{List}\;\Conid{Elm}\to \Conid{Bool}} can be defined by:
\begin{hscode}\SaveRestoreHook
\column{B}{@{}>{\hspre}l<{\hspost}@{}}%
\column{16}{@{}>{\hspre}l<{\hspost}@{}}%
\column{E}{@{}>{\hspre}l<{\hspost}@{}}%
\>[B]{}\Varid{sorted}\;[\mskip1.5mu \mskip1.5mu]{}\<[16]%
\>[16]{}\mathrel{=}\Conid{True}{}\<[E]%
\\
\>[B]{}\Varid{sorted}\;(\Varid{x}\mathbin{:}\Varid{xs}){}\<[16]%
\>[16]{}\mathrel{=}\Varid{all}\;(\Varid{x}\leq )\;\Varid{xs}\mathrel{\wedge}\Varid{sorted}\;\Varid{xs}~~.{}\<[E]%
\ColumnHook
\end{hscode}\resethooks
The following property can be proved by a routine induction on \ensuremath{\Varid{ys}}:
\begin{equation}
\begin{split}
&  \ensuremath{\Varid{sorted}\;(\Varid{ys}\mathbin{+\!\!\!+}[\mskip1.5mu \Varid{x}\mskip1.5mu]\mathbin{+\!\!\!+}\Varid{zs})} ~\equiv~ \\
&\quad    \ensuremath{\Varid{sorted}\;\Varid{ys}\mathrel{\wedge}\Varid{sorted}\;\Varid{zs}\mathrel{\wedge}\Varid{all}\;(\leq \Varid{x})\;\Varid{ys}\mathrel{\wedge}\Varid{all}\;(\Varid{x}\leq )\;\Varid{zs}}  \mbox{~~.}
\end{split}
    \label{eq:sorted-cat3}
\end{equation}

Now we consider the permutation phase.
As shown by Bird~\cite{Bird:96:Functional}, what sorting algorithm we end up deriving is often driven by how the permutation phase is performed.
The following definition of \ensuremath{\Varid{perm}}, for example:
\begin{hscode}\SaveRestoreHook
\column{B}{@{}>{\hspre}l<{\hspost}@{}}%
\column{14}{@{}>{\hspre}l<{\hspost}@{}}%
\column{E}{@{}>{\hspre}l<{\hspost}@{}}%
\>[B]{}\Varid{perm}\;[\mskip1.5mu \mskip1.5mu]{}\<[14]%
\>[14]{}\mathrel{=}\{[\mskip1.5mu \mskip1.5mu]\}{}\<[E]%
\\
\>[B]{}\Varid{perm}\;(\Varid{x}\mathbin{:}\Varid{xs}){}\<[14]%
\>[14]{}\mathrel{=}\Varid{perm}\;\Varid{xs}\mathrel{\hstretch{0.7}{>\!\!>\!\!=}}\Varid{insert}\;\Varid{x}~~,{}\<[E]%
\ColumnHook
\end{hscode}\resethooks
where \ensuremath{\Varid{insert}\;\Varid{x}\;\Varid{xs}} non-deterministically inserts \ensuremath{\Varid{x}} into \ensuremath{\Varid{xs}},
would lead us to insertion sort.
%
To derive quicksort, we use an alternative definition of \ensuremath{\Varid{perm}}:
\begin{hscode}\SaveRestoreHook
\column{B}{@{}>{\hspre}l<{\hspost}@{}}%
\column{14}{@{}>{\hspre}c<{\hspost}@{}}%
\column{14E}{@{}l@{}}%
\column{17}{@{}>{\hspre}l<{\hspost}@{}}%
\column{E}{@{}>{\hspre}l<{\hspost}@{}}%
\>[B]{}\Varid{perm}\mathbin{::}\Conid{MonadPlus}\;\Varid{m}\Rightarrow \Conid{List}\;\Varid{a}\to \Varid{m}\;(\Conid{List}\;\Varid{a}){}\<[E]%
\\
\>[B]{}\Varid{perm}\;[\mskip1.5mu \mskip1.5mu]{}\<[14]%
\>[14]{}\mathrel{=}{}\<[14E]%
\>[17]{}\{[\mskip1.5mu \mskip1.5mu]\}{}\<[E]%
\\
\>[B]{}\Varid{perm}\;(\Varid{x}\mathbin{:}\Varid{xs}){}\<[14]%
\>[14]{}\mathrel{=}{}\<[14E]%
\>[17]{}\Varid{split}\;\Varid{xs}\mathrel{\hstretch{0.7}{>\!\!>\!\!=}}\lambda (\Varid{ys},\Varid{zs})\to \Varid{liftM2}\;(\mathbin{+\!\!\!+}[\mskip1.5mu \Varid{x}\mskip1.5mu]\mathbin{+\!\!\!+})\;(\Varid{perm}\;\Varid{ys})\;(\Varid{perm}\;\Varid{zs})~~.{}\<[E]%
\ColumnHook
\end{hscode}\resethooks
where \ensuremath{\Varid{liftM2}\;(\oplus)\;\Varid{m}_{1}\;\Varid{m}_{2}\mathrel{=}\Varid{m}_{1}\mathrel{\hstretch{0.7}{>\!\!>\!\!=}}\lambda \Varid{x}_{1}\to \Varid{m}_{2}\mathrel{\hstretch{0.7}{>\!\!>\!\!=}}\lambda \Varid{x}_{2}\to \{\Varid{x}_{1}\mathbin{\oplus}\Varid{x}_{2}\}}, and \ensuremath{\Varid{split}} non-deterministically splits a list.
When the input has more than one element, we split the tail into two, permute them separately, and insert the head in the middle.
The monadic function \ensuremath{\Varid{split}} is given by:
\begin{hscode}\SaveRestoreHook
\column{B}{@{}>{\hspre}l<{\hspost}@{}}%
\column{15}{@{}>{\hspre}c<{\hspost}@{}}%
\column{15E}{@{}l@{}}%
\column{18}{@{}>{\hspre}l<{\hspost}@{}}%
\column{E}{@{}>{\hspre}l<{\hspost}@{}}%
\>[B]{}\Varid{split}\mathbin{::}\Conid{MonadPlus}\;\Varid{m}\Rightarrow \Conid{List}\;\Varid{a}\to \Varid{m}\;(\Conid{List}\;\Varid{a}\times\Conid{List}\;\Varid{a}){}\<[E]%
\\
\>[B]{}\Varid{split}\;[\mskip1.5mu \mskip1.5mu]{}\<[15]%
\>[15]{}\mathrel{=}{}\<[15E]%
\>[18]{}\{([\mskip1.5mu \mskip1.5mu],[\mskip1.5mu \mskip1.5mu])\}{}\<[E]%
\\
\>[B]{}\Varid{split}\;(\Varid{x}\mathbin{:}\Varid{xs}){}\<[15]%
\>[15]{}\mathrel{=}{}\<[15E]%
\>[18]{}\Varid{split}\;\Varid{xs}\mathrel{\hstretch{0.7}{>\!\!>\!\!=}}\lambda (\Varid{ys},\Varid{zs})\to \{(\Varid{x}\mathbin{:}\Varid{ys},\Varid{zs})\}\mathbin{\talloblong}\{(\Varid{ys},\Varid{x}\mathbin{:}\Varid{zs})\}~~.{}\<[E]%
\ColumnHook
\end{hscode}\resethooks
%
%
This completes the specification.
One may argue that the second definition of \ensuremath{\Varid{perm}} is not one that,
as stated in Section~\ref{sec:intro},
``obviously'' implied by the problem description.
Bird~\cite{Bird:96:Functional} derived the second one from the first in a relational setting, and we can also show that the two definitions are equivalent.



\section{Quicksort on Lists}
\label{sec:quicksort}

In this section we derive a divide-and-conquer property of \ensuremath{\Varid{slowsort}}.
It allows us to refine \ensuremath{\Varid{slowsort}} to the well-known recursive definition of quicksort on lists, and is also used in the next section to construct quicksort on arrays.

\subsubsection*{Refinement}
We will need to first define our concept of program refinement.
We abuse notations from set theory and define:
\begin{hscode}\SaveRestoreHook
\column{B}{@{}>{\hspre}l<{\hspost}@{}}%
\column{3}{@{}>{\hspre}l<{\hspost}@{}}%
\column{16}{@{}>{\hspre}l<{\hspost}@{}}%
\column{E}{@{}>{\hspre}l<{\hspost}@{}}%
\>[3]{}\Varid{m}_{1}\mathrel{\subseteq}\Varid{m}_{2}{}\<[16]%
\>[16]{}~\mathrel{\equiv}~\Varid{m}_{1}\mathbin{\talloblong}\Varid{m}_{2}\mathrel{=}\Varid{m}_{2}~~.{}\<[E]%
\ColumnHook
\end{hscode}\resethooks
The righthand side \ensuremath{\Varid{m}_{1}\mathbin{\talloblong}\Varid{m}_{2}\mathrel{=}\Varid{m}_{2}} says that every result of \ensuremath{\Varid{m}_{1}} is a possible result of \ensuremath{\Varid{m}_{2}}.
When \ensuremath{\Varid{m}_{1}\mathrel{\subseteq}\Varid{m}_{2}}, we say that \ensuremath{\Varid{m}_{1}} \emph{refines} \ensuremath{\Varid{m}_{1}}, \ensuremath{\Varid{m}_{2}} can be \emph{refined to} \ensuremath{\Varid{m}_{1}}, or that \ensuremath{\Varid{m}_{2}} \emph{subsumes} \ensuremath{\Varid{m}_{1}}.
Note that this definition applies not only to the non-determinism monad, but to monads having other effects as well.
We denote \ensuremath{(\subseteq)} lifted to functions by \ensuremath{(\dot{\subseteq})}:%
\begin{hscode}\SaveRestoreHook
\column{B}{@{}>{\hspre}l<{\hspost}@{}}%
\column{3}{@{}>{\hspre}l<{\hspost}@{}}%
\column{E}{@{}>{\hspre}l<{\hspost}@{}}%
\>[3]{}\Varid{f}\mathrel{\dot{\subseteq}}\Varid{g}~\mathrel{=}~(\forall \Varid{x}\hsforall \mathbin{:}\Varid{f}\;\Varid{x}\mathrel{\subseteq}\Varid{g}\;\Varid{x})~~.{}\<[E]%
\ColumnHook
\end{hscode}\resethooks
That is, \ensuremath{\Varid{f}} refines \ensuremath{\Varid{g}} if \ensuremath{\Varid{f}\;\Varid{x}} refines \ensuremath{\Varid{g}\;\Varid{x}} for all \ensuremath{\Varid{x}}.%
When we use this notation, \ensuremath{\Varid{f}} and \ensuremath{\Varid{g}} are always functions returning monads, which is sufficient for this pearl.

One can show that the definition of \ensuremath{(\subseteq)} is equivalent to
\ensuremath{\Varid{m}_{1}\mathrel{\subseteq}\Varid{m}_{2}\,\mathrel{\equiv}\,(\exists\;\Varid{n}\mathbin{:}\Varid{m}_{1}\mathbin{\talloblong}\Varid{n}\mathrel{=}\Varid{m}_{2})},
and that \ensuremath{(\subseteq)} and \ensuremath{(\dot{\subseteq})} are both reflexive, transitive, and anti-symmetric (\ensuremath{\Varid{m}\mathrel{\subseteq}\Varid{n}\mathrel{\wedge}\Varid{n}\mathrel{\subseteq}\Varid{m}\,\mathrel{\equiv}\,\Varid{n}\mathrel{=}\Varid{m}}). 
Furthermore, \ensuremath{(\mathrel{\hstretch{0.7}{>\!\!>\!\!=}})} respects refinement:
\begin{lemma}\label{lma:refine-bind-preservation}
Bind \ensuremath{(\mathrel{\hstretch{0.7}{>\!\!>\!\!=}})} is monotonic with respect to \ensuremath{(\subseteq)}.
That is,
\ensuremath{\Varid{m}_{1}\mathrel{\subseteq}\Varid{m}_{2}\,\mathrel{\Rightarrow}\,\Varid{m}_{1}\mathrel{\hstretch{0.7}{>\!\!>\!\!=}}\Varid{f}\mathrel{\subseteq}\Varid{m}_{2}\mathrel{\hstretch{0.7}{>\!\!>\!\!=}}\Varid{f}}, and
\ensuremath{\Varid{f}_{1}\mathrel{\dot{\subseteq}}\Varid{f}_{2}\,\mathrel{\Rightarrow}\,\Varid{m}\mathrel{\hstretch{0.7}{>\!\!>\!\!=}}\Varid{f}_{1}\mathrel{\subseteq}\Varid{m}\mathrel{\hstretch{0.7}{>\!\!>\!\!=}}\Varid{f}_{2}}.
\end{lemma}
Having Lemma~\ref{lma:refine-bind-preservation} allows us to refine programs in a compositional manner.
The proof of Lemma~\ref{lma:refine-bind-preservation} makes use of \eqref{eq:nd-left-distr} and \eqref{eq:nd-right-distr}.

\subsubsection*{Commutativity and \ensuremath{\Varid{guard}}}
We say that \ensuremath{\Varid{m}} and \ensuremath{\Varid{n}} commute if
\begin{hscode}\SaveRestoreHook
\column{B}{@{}>{\hspre}l<{\hspost}@{}}%
\column{E}{@{}>{\hspre}l<{\hspost}@{}}%
\>[B]{}\Varid{m}\mathrel{\hstretch{0.7}{>\!\!>\!\!=}}\lambda \Varid{x}\to \Varid{n}\mathrel{\hstretch{0.7}{>\!\!>\!\!=}}\lambda \Varid{y}\to \Varid{f}\;\Varid{x}\;\Varid{y}~\mathrel{=}~\Varid{n}\mathrel{\hstretch{0.7}{>\!\!>\!\!=}}\lambda \Varid{y}\to \Varid{m}\mathrel{\hstretch{0.7}{>\!\!>\!\!=}}\lambda \Varid{x}\to \Varid{f}\;\Varid{x}\;\Varid{y}~~.{}\<[E]%
\ColumnHook
\end{hscode}\resethooks
It can be proved that \ensuremath{\Varid{guard}\;\Varid{p}} commutes with all \ensuremath{\Varid{m}} if non-determinism is the only effect in \ensuremath{\Varid{m}} --- a property we will need many times.
Furthermore, having right-zero \eqref{eq:nd-right-zero} and right-distributivity \eqref{eq:nd-right-distr}, in addition to other laws, one can prove that non-determinism commutes with other effects.
In particular, non-determinism commutes with state.

We mention two more properties about \ensuremath{\Varid{guard}}:
\ensuremath{\Varid{guard}\;(\Varid{p}\mathrel{\wedge}\Varid{q})} can be split into two, and \ensuremath{\Varid{guard}}s with complementary predicates can be refined to \ensuremath{\mathbf{if}}:
\begin{align}
\ensuremath{\Varid{guard}\;(\Varid{p}\mathrel{\wedge}\Varid{q})} &~= \ensuremath{\Varid{guard}\;\Varid{p}\mathbin{\hstretch{0.7}{>\!\!>}}\Varid{guard}\;\Varid{q}} \mbox{~~,}
\label{eq:guard-conj-split}\\
\ensuremath{(\Varid{guard}\;\Varid{p}\mathbin{\hstretch{0.7}{>\!\!>}}\Varid{m}_{1})} &\ensuremath{\mathbin{\talloblong}(\Varid{guard}\;(\neg \mathbin{\cdot}\Varid{p})\mathbin{\hstretch{0.7}{>\!\!>}}\Varid{m}_{2})} ~\ensuremath{\mathrel{\supseteq}~\;\mathbf{if}\;\Varid{p}\;\mathbf{then}\;\Varid{m}_{1}\;\mathbf{else}\;\Varid{m}_{2}} \mbox{~~.}
\label{eq:guard-if}
\end{align}

\subsubsection*{Divide-and-Conquer}
Back to \ensuremath{\Varid{slowsort}}. We proceed with usual routine in functional programming: case-analysis on the input.
For the base case, \ensuremath{\Varid{slowsort}\;[\mskip1.5mu \mskip1.5mu]\mathrel{=}\{[\mskip1.5mu \mskip1.5mu]\}}.
For the inductive case, the crucial step is the commutativity of \ensuremath{\Varid{guard}}:
\begin{hscode}\SaveRestoreHook
\column{B}{@{}>{\hspre}l<{\hspost}@{}}%
\column{7}{@{}>{\hspre}l<{\hspost}@{}}%
\column{9}{@{}>{\hspre}l<{\hspost}@{}}%
\column{E}{@{}>{\hspre}l<{\hspost}@{}}%
\>[7]{}\Varid{slowsort}\;(\Varid{p}\mathbin{:}\Varid{xs}){}\<[E]%
\\
\>[B]{}\mathbin{=}{}\<[9]%
\>[9]{}\mbox{\commentbegin  expanding definitions, monad laws  \commentend}{}\<[E]%
\\
\>[B]{}\hsindent{7}{}\<[7]%
\>[7]{}\Varid{split}\;\Varid{xs}\mathrel{\hstretch{0.7}{>\!\!>\!\!=}}\lambda (\Varid{ys},\Varid{zs})\to {}\<[E]%
\\
\>[B]{}\hsindent{7}{}\<[7]%
\>[7]{}\Varid{perm}\;\Varid{ys}\mathrel{\hstretch{0.7}{>\!\!>\!\!=}}\lambda \Varid{ys'}\to \Varid{perm}\;\Varid{zs}\mathrel{\hstretch{0.7}{>\!\!>\!\!=}}\lambda \Varid{zs'}\to {}\<[E]%
\\
\>[B]{}\hsindent{7}{}\<[7]%
\>[7]{}\Varid{filt}\;\Varid{sorted}\;(\Varid{ys'}\mathbin{+\!\!\!+}[\mskip1.5mu \Varid{p}\mskip1.5mu]\mathbin{+\!\!\!+}\Varid{zs'}){}\<[E]%
\\
\>[B]{}\mathbin{=}{}\<[9]%
\>[9]{}\mbox{\commentbegin  by \eqref{eq:sorted-cat3}  \commentend}{}\<[E]%
\\
\>[B]{}\hsindent{7}{}\<[7]%
\>[7]{}\Varid{split}\;\Varid{xs}\mathrel{\hstretch{0.7}{>\!\!>\!\!=}}\lambda (\Varid{ys},\Varid{zs})\to {}\<[E]%
\\
\>[B]{}\hsindent{7}{}\<[7]%
\>[7]{}\Varid{perm}\;\Varid{ys}\mathrel{\hstretch{0.7}{>\!\!>\!\!=}}\lambda \Varid{ys'}\to \Varid{perm}\;\Varid{zs}\mathrel{\hstretch{0.7}{>\!\!>\!\!=}}\lambda \Varid{zs'}\to {}\<[E]%
\\
\>[B]{}\hsindent{7}{}\<[7]%
\>[7]{}\Varid{guard}\;(\Varid{sorted}\;\Varid{ys'}\mathrel{\wedge}\Varid{sorted}\;\Varid{zs'}\mathrel{\wedge}\Varid{all}\;(\leq \Varid{p})\;\Varid{ys'}\mathrel{\wedge}\Varid{all}\;(\Varid{p}\leq )\;\Varid{zs'})\mathbin{\hstretch{0.7}{>\!\!>}}{}\<[E]%
\\
\>[B]{}\hsindent{7}{}\<[7]%
\>[7]{}\{\Varid{ys'}\mathbin{+\!\!\!+}[\mskip1.5mu \Varid{p}\mskip1.5mu]\mathbin{+\!\!\!+}\Varid{zs'}\}{}\<[E]%
\\
\>[B]{}\mathbin{=}{}\<[9]%
\>[9]{}\mbox{\commentbegin  \eqref{eq:guard-conj-split} and that \ensuremath{\Varid{guard}} commutes with non-determinism  \commentend}{}\<[E]%
\\
\>[B]{}\hsindent{7}{}\<[7]%
\>[7]{}\Varid{split}\;\Varid{xs}\mathrel{\hstretch{0.7}{>\!\!>\!\!=}}\lambda (\Varid{ys},\Varid{zs})\to \Varid{guard}\;(\Varid{all}\;(\leq \Varid{p})\;\Varid{ys}\mathrel{\wedge}\Varid{all}\;(\Varid{p}\leq )\;\Varid{zs'})\mathbin{\hstretch{0.7}{>\!\!>}}{}\<[E]%
\\
\>[B]{}\hsindent{7}{}\<[7]%
\>[7]{}(\Varid{perm}\;\Varid{ys}\mathrel{\hstretch{0.7}{>\!\!>\!\!=}}\Varid{filt}\;\Varid{sorted})\mathrel{\hstretch{0.7}{>\!\!>\!\!=}}\lambda \Varid{ys'}\to {}\<[E]%
\\
\>[B]{}\hsindent{7}{}\<[7]%
\>[7]{}(\Varid{perm}\;\Varid{zs}\mathrel{\hstretch{0.7}{>\!\!>\!\!=}}\Varid{filt}\;\Varid{sorted})\mathrel{\hstretch{0.7}{>\!\!>\!\!=}}\lambda \Varid{zs'}\to {}\<[E]%
\\
\>[B]{}\hsindent{7}{}\<[7]%
\>[7]{}\{\Varid{ys'}\mathbin{+\!\!\!+}[\mskip1.5mu \Varid{p}\mskip1.5mu]\mathbin{+\!\!\!+}\Varid{zs'}\}~~.{}\<[E]%
\ColumnHook
\end{hscode}\resethooks
Provided that we can construct a function \ensuremath{\Varid{partition}} such that
\begin{hscode}\SaveRestoreHook
\column{B}{@{}>{\hspre}l<{\hspost}@{}}%
\column{E}{@{}>{\hspre}l<{\hspost}@{}}%
\>[B]{}\{\Varid{partition}\;\Varid{p}\;\Varid{xs}\}~\mathrel{\subseteq}~\Varid{split}\;\Varid{xs}\mathrel{\hstretch{0.7}{>\!\!>\!\!=}}\Varid{filt}\;(\lambda (\Varid{ys},\Varid{zs})\to \Varid{all}\;(\leq \Varid{p})\;\Varid{ys}\mathrel{\wedge}\Varid{all}\;(\Varid{p}\leq )\;\Varid{zs})~~,{}\<[E]%
\ColumnHook
\end{hscode}\resethooks
we have established the following divide-and-conquer property:
\begin{equation}
\begin{split}
\ensuremath{\Varid{slowsort}\;(\Varid{p}\mathbin{:}\Varid{xs})} ~~\supseteq~~ & \ensuremath{\{\Varid{partition}\;\Varid{p}\;\Varid{xs}\}\mathrel{\hstretch{0.7}{>\!\!>\!\!=}}\lambda (\Varid{ys},\Varid{zs})\to }\\[-1mm]
                       & \ensuremath{\Varid{slowsort}\;\Varid{ys}\mathrel{\hstretch{0.7}{>\!\!>\!\!=}}\lambda \Varid{ys'}\to \Varid{slowsort}\;\Varid{zs}\mathrel{\hstretch{0.7}{>\!\!>\!\!=}}\lambda \Varid{zs'}\to }\\[-1mm]
                       & \ensuremath{\{\Varid{ys'}\mathbin{+\!\!\!+}[\mskip1.5mu \Varid{p}\mskip1.5mu]\mathbin{+\!\!\!+}\Varid{zs'}\}~~.}
\end{split}
\label{eq:slowsort-rec}
\end{equation}


The derivation of \ensuremath{\Varid{partition}} proceeds by induction on the input.
In the case for \ensuremath{\Varid{xs}\mathbin{:=}\Varid{x}\mathbin{:}\Varid{xs}} we need to refine two guarded choices, \ensuremath{(\Varid{guard}\;(\Varid{x}\leq \Varid{p})\mathbin{\hstretch{0.7}{>\!\!>}}\{\Varid{x}\mathbin{:}\Varid{ys},\Varid{zs}\})\mathbin{\talloblong}(\Varid{guard}\;(\Varid{p}\leq \Varid{x})\mathbin{\hstretch{0.7}{>\!\!>}}\{\Varid{ys},\Varid{x}\mathbin{:}\Varid{zs}\})}, to an \ensuremath{\mathbf{if}} branching. When \ensuremath{\Varid{x}} and \ensuremath{\Varid{p}} equal, the specification allows us to place \ensuremath{\Varid{x}} in either partition. For no particular reason, we choose the left partition. That gives us:
\begin{hscode}\SaveRestoreHook
\column{B}{@{}>{\hspre}l<{\hspost}@{}}%
\column{21}{@{}>{\hspre}c<{\hspost}@{}}%
\column{21E}{@{}l@{}}%
\column{24}{@{}>{\hspre}l<{\hspost}@{}}%
\column{E}{@{}>{\hspre}l<{\hspost}@{}}%
\>[B]{}\Varid{partition}\;\Varid{p}\;[\mskip1.5mu \mskip1.5mu]{}\<[21]%
\>[21]{}\mathrel{=}{}\<[21E]%
\>[24]{}([\mskip1.5mu \mskip1.5mu],[\mskip1.5mu \mskip1.5mu]){}\<[E]%
\\
\>[B]{}\Varid{partition}\;\Varid{p}\;(\Varid{x}\mathbin{:}\Varid{xs}){}\<[21]%
\>[21]{}\mathrel{=}{}\<[21E]%
\>[24]{}\mathbf{let}\;(\Varid{ys},\Varid{zs})\mathrel{=}\Varid{partition}\;\Varid{p}\;\Varid{xs}{}\<[E]%
\\
\>[24]{}\mathbf{in}\;\mathbf{if}\;\Varid{x}\leq \Varid{p}\;\mathbf{then}\;(\Varid{x}\mathbin{:}\Varid{ys},\Varid{zs})\;\mathbf{else}\;(\Varid{ys},\Varid{x}\mathbin{:}\Varid{zs})~~.{}\<[E]%
\ColumnHook
\end{hscode}\resethooks
Having \ensuremath{\Varid{partition}} derived, it takes only a routine induction on the length of input lists to show that \ensuremath{\{\cdot \}\mathbin{\cdot}\Varid{qsort}\mathrel{\dot{\subseteq}}\Varid{slowsort}}, where \ensuremath{\Varid{qsort}} is given by:
\begin{hscode}\SaveRestoreHook
\column{B}{@{}>{\hspre}l<{\hspost}@{}}%
\column{15}{@{}>{\hspre}c<{\hspost}@{}}%
\column{15E}{@{}l@{}}%
\column{18}{@{}>{\hspre}l<{\hspost}@{}}%
\column{E}{@{}>{\hspre}l<{\hspost}@{}}%
\>[B]{}\Varid{qsort}\;[\mskip1.5mu \mskip1.5mu]{}\<[15]%
\>[15]{}\mathrel{=}{}\<[15E]%
\>[18]{}[\mskip1.5mu \mskip1.5mu]{}\<[E]%
\\
\>[B]{}\Varid{qsort}\;(\Varid{p}\mathbin{:}\Varid{xs}){}\<[15]%
\>[15]{}\mathrel{=}{}\<[15E]%
\>[18]{}\mathbf{let}\;(\Varid{ys},\Varid{zs})\mathrel{=}\Varid{partition}\;\Varid{p}\;\Varid{xs}{}\<[E]%
\\
\>[18]{}\mathbf{in}\;\Varid{qsort}\;\Varid{ys}\mathbin{+\!\!\!+}[\mskip1.5mu \Varid{p}\mskip1.5mu]\mathbin{+\!\!\!+}\Varid{qsort}\;\Varid{zs}~~.{}\<[E]%
\ColumnHook
\end{hscode}\resethooks
As is typical in program derivation, the termination of derived program is shown separately afterwards. In this case, \ensuremath{\Varid{qsort}} terminates because the input list decreases in size in every recursive call --- for that we need to show that, in the call to \ensuremath{\Varid{partition}}, the sum of lengths of \ensuremath{\Varid{ys}} and \ensuremath{\Varid{zs}} equals that of \ensuremath{\Varid{xs}}.

\section{Quicksort on Arrays}
\label{sec:iqsort}

One of the advantages of using a monadic calculus is that we can integrate effects other than non-determinism into the program we derive.
In this section we derive an imperative quicksort on arrays, based on previously established properties.



\subsection{Operations on Arrays}
\label{sec:array-operations}

We assume that our state is an \ensuremath{\Conid{Int}}-indexed, unbounded array containing elements of type \ensuremath{\Varid{e}}, with two operations that, given an index, respectively read from and write to the array:
\begin{hscode}\SaveRestoreHook
\column{B}{@{}>{\hspre}l<{\hspost}@{}}%
\column{3}{@{}>{\hspre}l<{\hspost}@{}}%
\column{10}{@{}>{\hspre}l<{\hspost}@{}}%
\column{E}{@{}>{\hspre}l<{\hspost}@{}}%
\>[B]{}\mathbf{class}\;\Conid{Monad}\;\Varid{m}\Rightarrow \Conid{MonadArr}\;\Varid{e}\;\Varid{m}\;\mathbf{where}{}\<[E]%
\\
\>[B]{}\hsindent{3}{}\<[3]%
\>[3]{}\Varid{read}{}\<[10]%
\>[10]{}\mathbin{::}\Conid{Int}\to \Varid{m}\;\Varid{e}{}\<[E]%
\\
\>[B]{}\hsindent{3}{}\<[3]%
\>[3]{}\Varid{write}{}\<[10]%
\>[10]{}\mathbin{::}\Conid{Int}\to \Varid{e}\to \Varid{m}\;()~~.{}\<[E]%
\ColumnHook
\end{hscode}\resethooks
They are assumed to satisfy the following laws:
\begin{align*}
  &\mbox{\bf read-write:} &
  \ensuremath{\Varid{read}\;\Varid{i}\mathrel{\hstretch{0.7}{>\!\!>\!\!=}}\Varid{write}\;\Varid{i}} ~&=~ \ensuremath{\{()\}} \mbox{~~,}\\
  &\mbox{\bf write-read:} &
  \ensuremath{\Varid{write}\;\Varid{i}\;\Varid{x}\mathbin{\hstretch{0.7}{>\!\!>}}\Varid{read}\;\Varid{i}} ~&=~
  \ensuremath{\Varid{write}\;\Varid{i}\;\Varid{x}\mathbin{\hstretch{0.7}{>\!\!>}}\{\Varid{x}\}} \mbox{~~,}\\
  &\mbox{\bf write-write:}&
  \ensuremath{\Varid{write}\;\Varid{i}\;\Varid{x}\mathbin{\hstretch{0.7}{>\!\!>}}\Varid{write}\;\Varid{i}\;\Varid{x'}} ~&=~ \ensuremath{\Varid{write}\;\Varid{i}\;\Varid{x'}} \mbox{~~,}\\
  &\mbox{\bf read-read:}  &
  \multispan2{\ensuremath{\Varid{read}\;\Varid{i}\mathrel{\hstretch{0.7}{>\!\!>\!\!=}}\lambda \Varid{x}\to \Varid{read}\;\Varid{i}\mathrel{\hstretch{0.7}{>\!\!>\!\!=}}\lambda \Varid{x'}\to \Varid{f}\;\Varid{x}\;\Varid{x'}}~=}\\
  && \multispan2{\qquad\ensuremath{\Varid{read}\;\Varid{i}\mathrel{\hstretch{0.7}{>\!\!>\!\!=}}\lambda \Varid{x}\to \Varid{f}\;\Varid{x}\;\Varid{x}} \mbox{~~.}\hfil}
\end{align*}
Furthermore, we assume that (1) \ensuremath{\Varid{read}\;\Varid{i}} and \ensuremath{\Varid{read}\;\Varid{j}} commute; (2) \ensuremath{\Varid{write}\;\Varid{i}\;\Varid{x}} and \ensuremath{\Varid{write}\;\Varid{j}\;\Varid{y}} commute if \ensuremath{\Varid{i}\mathbin{\not\doubleequals}\Varid{j}}; (3) \ensuremath{\Varid{write}\;\Varid{i}\;\Varid{x}} and \ensuremath{\Varid{read}\;\Varid{j}} commute if \ensuremath{\Varid{i}\mathbin{\not\doubleequals}\Varid{j}}.

\begin{figure}[t]
\hspace{-1cm}
{\small
\begin{hscode}\SaveRestoreHook
\column{B}{@{}>{\hspre}l<{\hspost}@{}}%
\column{11}{@{}>{\hspre}l<{\hspost}@{}}%
\column{19}{@{}>{\hspre}l<{\hspost}@{}}%
\column{21}{@{}>{\hspre}l<{\hspost}@{}}%
\column{25}{@{}>{\hspre}l<{\hspost}@{}}%
\column{E}{@{}>{\hspre}l<{\hspost}@{}}%
\>[B]{}\Varid{readList}\mathbin{::}\Conid{MonadArr}\;\Varid{e}\;\Varid{m}\Rightarrow \Conid{Int}\to \Conid{Nat}\to \Varid{m}\;(\Conid{List}\;\Varid{e}){}\<[E]%
\\
\>[B]{}\Varid{readList}\;\Varid{i}\;\mathrm{0}{}\<[19]%
\>[19]{}\mathrel{=}\{[\mskip1.5mu \mskip1.5mu]\}{}\<[E]%
\\
\>[B]{}\Varid{readList}\;\Varid{i}\;(\mathrm{1}\mathbin{+}\Varid{k}){}\<[19]%
\>[19]{}\mathrel{=}\Varid{liftM2}\;(\mathbin{:})\;(\Varid{read}\;\Varid{i})\;(\Varid{readList}\;(\Varid{i}\mathbin{+}\mathrm{1})\;\Varid{k})~~,{}\<[E]%
\\[\blanklineskip]%
\>[B]{}\Varid{writeList}\mathbin{::}\Conid{MonadArr}\;\Varid{e}\;\Varid{m}\Rightarrow \Conid{Int}\to \Conid{List}\;\Varid{e}\to \Varid{m}\;(){}\<[E]%
\\
\>[B]{}\Varid{writeList}\;\Varid{i}\;[\mskip1.5mu \mskip1.5mu]{}\<[21]%
\>[21]{}\mathrel{=}\{()\}{}\<[E]%
\\
\>[B]{}\Varid{writeList}\;\Varid{i}\;(\Varid{x}\mathbin{:}\Varid{xs}){}\<[21]%
\>[21]{}\mathrel{=}\Varid{write}\;\Varid{i}\;\Varid{x}\mathbin{\hstretch{0.7}{>\!\!>}}\Varid{writeList}\;(\Varid{i}\mathbin{+}\mathrm{1})\;\Varid{xs}~~,{}\<[E]%
\\[\blanklineskip]%
\>[B]{}\Varid{writeL}\;{}\<[11]%
\>[11]{}\Varid{i}\;\Varid{xs}{}\<[25]%
\>[25]{}\mathrel{=}\Varid{writeList}\;\Varid{i}\;\Varid{xs}\mathbin{\hstretch{0.7}{>\!\!>}}\{\#\!\!\;\Varid{xs}\}~~,{}\<[E]%
\\
\>[B]{}\Varid{write2L}\;{}\<[11]%
\>[11]{}\Varid{i}\;(\Varid{xs},\Varid{ys}){}\<[25]%
\>[25]{}\mathrel{=}\Varid{writeList}\;\Varid{i}\;(\Varid{xs}\mathbin{+\!\!\!+}\Varid{ys})\mathbin{\hstretch{0.7}{>\!\!>}}\{(\#\!\!\;\Varid{xs},\#\!\!\;\Varid{ys})\}~~,{}\<[E]%
\\
\>[B]{}\Varid{write3L}\;{}\<[11]%
\>[11]{}\Varid{i}\;(\Varid{xs},\Varid{ys},\Varid{zs}){}\<[25]%
\>[25]{}\mathrel{=}\Varid{writeList}\;\Varid{i}\;(\Varid{xs}\mathbin{+\!\!\!+}\Varid{ys}\mathbin{+\!\!\!+}\Varid{zs})\mathbin{\hstretch{0.7}{>\!\!>}}\{(\#\!\!\;\Varid{xs},\#\!\!\;\Varid{ys},\#\!\!\;\Varid{zs})\}~~.{}\<[E]%
\\[\blanklineskip]%
\>[B]{}\Varid{swap}\;\Varid{i}\;\Varid{j}\mathrel{=}\Varid{read}\;\Varid{i}\mathrel{\hstretch{0.7}{>\!\!>\!\!=}}\lambda \Varid{x}\to \Varid{read}\;\Varid{j}\mathrel{\hstretch{0.7}{>\!\!>\!\!=}}\lambda \Varid{y}\to \Varid{write}\;\Varid{i}\;\Varid{y}\mathbin{\hstretch{0.7}{>\!\!>}}\Varid{write}\;\Varid{j}\;\Varid{x}~~.{}\<[E]%
\ColumnHook
\end{hscode}\resethooks
} 
\vspace{-0.7cm}
\caption{Operations for reading and writing chunks of data.}
\label{fig:readWriteList}
\end{figure}
%
More operations defined in terms of \ensuremath{\Varid{read}} and \ensuremath{\Varid{write}} are shown in Figure~\ref{fig:readWriteList}, where \ensuremath{\#\!\!\;\Varid{xs}} abbreviates \ensuremath{\Varid{length}~\Varid{xs}}.
The function \ensuremath{\Varid{readList}\;\Varid{i}\;\Varid{n}}, where \ensuremath{\Varid{n}} is a natural number, returns a list containing the \ensuremath{\Varid{n}} elements in the array starting from index \ensuremath{\Varid{i}}. Conversely, \ensuremath{\Varid{writeList}\;\Varid{i}\;\Varid{xs}} writes the list \ensuremath{\Varid{xs}} to the array with the first element being at index \ensuremath{\Varid{i}}.
In imperative programming we often store sequences of data into an array and return the length of the data. Thus, functions
\ensuremath{\Varid{writeL}}, \ensuremath{\Varid{write2L}} and \ensuremath{\Varid{write3L}} store lists into the array before returning their lengths.
These \ensuremath{\Varid{read}} and \ensuremath{\Varid{write}} family of functions are used only in the specification; the algorithm we construct should only mutate the array by \ensuremath{\Varid{swap}}ing elements.

Among the many properties of \ensuremath{\Varid{readList}} and \ensuremath{\Varid{writeList}} that can be induced from their definitions, the following will be used in a number of crucial steps:
\begin{align}
  \ensuremath{\Varid{writeList}\;\Varid{i}\;(\Varid{xs}\mathbin{+\!\!\!+}\Varid{ys})} ~=~ \ensuremath{\Varid{writeList}\;\Varid{i}\;\Varid{xs}\mathbin{\hstretch{0.7}{>\!\!>}}\Varid{writeList}\;(\Varid{i}\mathbin{+}\#\!\!\;\Varid{xs})\;\Varid{ys}} \mbox{~~.} \label{eq:writeList-++}
\end{align}

A function \ensuremath{\Varid{f}\mathbin{::}\Conid{List}\;\Varid{a}\to \Varid{m}\;(\Conid{List}\;\Varid{a})} is said to be \emph{length preserving} if
\ensuremath{\Varid{f}\;\Varid{xs}\mathrel{\hstretch{0.7}{>\!\!>\!\!=}}\lambda \Varid{ys}\to \{(\Varid{ys},\#\!\!\;\Varid{ys})\}} \ensuremath{\mathrel{=}}
\ensuremath{\Varid{f}\;\Varid{xs}\mathrel{\hstretch{0.7}{>\!\!>\!\!=}}\lambda \Varid{ys}\to \{(\Varid{ys},\#\!\!\;\Varid{xs})\}}.
It can be proved that \ensuremath{\Varid{perm}}, and thus \ensuremath{\Varid{slowsort}}, are length preserving.

\paragraph{On ``composing monads''}
In the sections to follow, some readers may have concern seeing \ensuremath{\Varid{perm}}, having class constraint \ensuremath{\Conid{MonadPlus}\;\Varid{m}}, and some other code having constraint \ensuremath{\Conid{MonadArr}\;\Varid{e}\;\Varid{m}} in the same expression.
This is totally fine:
mixing two such subterms simply results in an expression having constraint \ensuremath{(\Conid{MonadPlus}\;\Varid{m},\Conid{MonadArr}\;\Varid{e}\;\Varid{m})}.
No \ensuremath{\Varid{lift}}ing is necessary.

We use type classes to make it clear that we do not specify what exact monad \ensuremath{\Varid{perm}} is implemented with.
It could be one monolithic monad, a monad built from monad transformers~\cite{MTL:14}, or a free monad interpreted by effect handlers~\cite{KiselyovIshii:15:Freer}.
All theorems and derivations about \ensuremath{\Varid{perm}} hold regardless of the actual monad, as long as the monad satisfies all properties we demand.

\subsection{Partitioning an Array}

While the list-based \ensuremath{\Varid{partition}} is relatively intuitive,
partitioning an array \emph{in-place} (that is, using at most $O(1)$ additional space) is known to be a tricky phase of array-based quicksort.
Therefore we commence our discussion from deriving in-place array partitioning  from the list version.
The partition algorithm we end up deriving is known as the \emph{Lomuto scheme}~\cite{Bentley:00:Programming}, as opposed to Hoare's~\cite{Hoare:61:Partition}.

\subsubsection*{Specification}
There are two issues to deal with before we present a specification for an imperative, array-based partitioning, based on list-based \ensuremath{\Varid{partition}}.
Firstly, \ensuremath{\Varid{partition}} is not tail-recursive, while many linear-time array algorithms are implemented as a tail-recursive $\Conid{for}$-loop.
Thus we apply the standard trick constructing a tail-recursive algorithm by introducing accumulating parameters.
Define (we write the input/outputs of \ensuremath{\Varid{partition}} in bold font for clarity):
\begin{hscode}\SaveRestoreHook
\column{B}{@{}>{\hspre}l<{\hspost}@{}}%
\column{30}{@{}>{\hspre}l<{\hspost}@{}}%
\column{E}{@{}>{\hspre}l<{\hspost}@{}}%
\>[B]{}\Varid{partl}\mathbin{::}\Conid{Elm}\to (\Conid{List}\;\Conid{Elm}\times\Conid{List}\;\Conid{Elm}\times\Conid{List}\;\Conid{Elm})\to (\Conid{List}\;\Conid{Elm}\times\Conid{List}\;\Conid{Elm}){}\<[E]%
\\
\>[B]{}\Varid{partl}\;\Varid{p}\;(\Varid{ys},\Varid{zs},\boldsymbol{\Varid{xs}})\mathrel{=}{}\<[30]%
\>[30]{}\mathbf{let}\;(\boldsymbol{\Varid{us}},\boldsymbol{\Varid{vs}})\mathrel{=}\Varid{partition}\;\Varid{p}\;\boldsymbol{\Varid{xs}}{}\<[E]%
\\
\>[30]{}\mathbf{in}\;(\Varid{ys}\mathbin{+\!\!\!+}\boldsymbol{\Varid{us}},\Varid{zs}\mathbin{+\!\!\!+}\boldsymbol{\Varid{vs}})~~.{}\<[E]%
\ColumnHook
\end{hscode}\resethooks
In words, \ensuremath{\Varid{partl}\;\Varid{p}\;(\Varid{ys},\Varid{zs},\boldsymbol{\Varid{xs}})} partitions \ensuremath{\boldsymbol{\Varid{xs}}} into \ensuremath{(\boldsymbol{\Varid{us}},\boldsymbol{\Varid{vs}})} with respect to pivot \ensuremath{\Varid{p}}, but appends \ensuremath{\Varid{ys}} and \ensuremath{\Varid{zs}} respectively to \ensuremath{\boldsymbol{\Varid{us}}} and \ensuremath{\boldsymbol{\Varid{vs}}}.
It is a generalisation of \ensuremath{\Varid{partition}} because \ensuremath{\Varid{partition}\;\Varid{p}\;\Varid{xs}\mathrel{=}\Varid{partl}\;\Varid{p}\;([\mskip1.5mu \mskip1.5mu],[\mskip1.5mu \mskip1.5mu],\Varid{xs})}.
By routine calculation exploiting associativity of \ensuremath{(\mathbin{+\!\!\!+})}, we can derive a tail-recursive definition of \ensuremath{\Varid{partl}}:
\begin{hscode}\SaveRestoreHook
\column{B}{@{}>{\hspre}l<{\hspost}@{}}%
\column{27}{@{}>{\hspre}l<{\hspost}@{}}%
\column{41}{@{}>{\hspre}l<{\hspost}@{}}%
\column{47}{@{}>{\hspre}l<{\hspost}@{}}%
\column{E}{@{}>{\hspre}l<{\hspost}@{}}%
\>[B]{}\Varid{partl}\;\Varid{p}\;(\Varid{ys},\Varid{zs},[\mskip1.5mu \mskip1.5mu]){}\<[27]%
\>[27]{}\mathrel{=}(\Varid{ys},\Varid{zs}){}\<[E]%
\\
\>[B]{}\Varid{partl}\;\Varid{p}\;(\Varid{ys},\Varid{zs},\bm{\Varid{x}}\mathbin{:}\bm{\Varid{xs}}){}\<[27]%
\>[27]{}\mathrel{=}\mathbf{if}\;\bm{\Varid{x}}\leq \Varid{p}\;{}\<[41]%
\>[41]{}\mathbf{then}\;{}\<[47]%
\>[47]{}\Varid{partl}\;\Varid{p}\;(\Varid{ys}\mathbin{+\!\!\!+}[\mskip1.5mu \bm{\Varid{x}}\mskip1.5mu],\Varid{zs},\bm{\Varid{xs}}){}\<[E]%
\\
\>[41]{}\mathbf{else}\;{}\<[47]%
\>[47]{}\Varid{partl}\;\Varid{p}\;(\Varid{ys},\Varid{zs}\mathbin{+\!\!\!+}[\mskip1.5mu \bm{\Varid{x}}\mskip1.5mu],\bm{\Varid{xs}})~~.{}\<[E]%
\ColumnHook
\end{hscode}\resethooks
It might aid our understanding if we note that, if we start \ensuremath{\Varid{partl}} with initial value \ensuremath{([\mskip1.5mu \mskip1.5mu],[\mskip1.5mu \mskip1.5mu],\Varid{xs})} we have the invariant that \ensuremath{\Varid{ys}} contains elements that are at most \ensuremath{\Varid{p}}, and elements in \ensuremath{\Varid{zs}} are larger than \ensuremath{\Varid{p}}. The calculations below, however, do not rely on this observation.
\footnote{It might be worth noting that \ensuremath{\Varid{partl}} causes a space
leak in Haskell, since the accumulators become thunks
that increase in size as the input list is traversed.
It does not matter here since \ensuremath{\Varid{partl}} merely serves as a specification of \ensuremath{\Varid{ipartl}}.}

Our wish is to construct a variant of \ensuremath{\Varid{partl}} that works on arrays.
That is, when the array contains \ensuremath{\Varid{ys}\mathbin{+\!\!\!+}\Varid{zs}\mathbin{+\!\!\!+}\boldsymbol{\Varid{xs}}}, the three inputs to \ensuremath{\Varid{partl}} in a consecutive segment, when the derived program finishes its work we wish to have \ensuremath{\Varid{ys}\mathbin{+\!\!\!+}\boldsymbol{\Varid{us}}\mathbin{+\!\!\!+}\Varid{zs}\mathbin{+\!\!\!+}\boldsymbol{\Varid{vs}}}, the output of \ensuremath{\Varid{partl}}, stored consecutively in the array.

This brings us to the second issue: \ensuremath{\Varid{partition}}, and therefore \ensuremath{\Varid{partl}}, are stable (that is, elements in each partition retain their original order), which is a strong requirement for array-based partitioning.
It is costly to mutate \ensuremath{\Varid{ys}\mathbin{+\!\!\!+}\Varid{zs}\mathbin{+\!\!\!+}\boldsymbol{\Varid{xs}}} into \ensuremath{\Varid{ys}\mathbin{+\!\!\!+}\boldsymbol{\Varid{us}}\mathbin{+\!\!\!+}\Varid{zs}\mathbin{+\!\!\!+}\boldsymbol{\Varid{vs}}}, since it demands that we retain the order of elements in \ensuremath{\Varid{zs}} while inserting elements of \ensuremath{\boldsymbol{\Varid{us}}}.
For sorting we do not need such a strong postcondition.
It is sufficient, and can be done more efficiently, to mutate \ensuremath{\Varid{ys}\mathbin{+\!\!\!+}\Varid{zs}\mathbin{+\!\!\!+}\boldsymbol{\Varid{xs}}} into \ensuremath{\Varid{ys}\mathbin{+\!\!\!+}\boldsymbol{\Varid{us}}\mathbin{+\!\!\!+}\Varid{ws}}, where \ensuremath{\Varid{ws}} is some permutation of \ensuremath{\Varid{zs}\mathbin{+\!\!\!+}\boldsymbol{\Varid{vs}}}.
It is handy allowing non-determinism: we introduce a \ensuremath{\Varid{perm}} in our specification, indicating that we do not care about the order of elements in \ensuremath{\Varid{ws}}.

Define \ensuremath{\Varid{second}\mathbin{::}\Conid{Monad}\;\Varid{m}\Rightarrow (\Varid{b}\to \Varid{m}\;\Varid{c})\to (\Varid{a},\Varid{b})\to \Varid{m}\;(\Varid{a},\Varid{c})},
which applies a monadic function to the second component of a tuple:
\begin{hscode}\SaveRestoreHook
\column{B}{@{}>{\hspre}l<{\hspost}@{}}%
\column{E}{@{}>{\hspre}l<{\hspost}@{}}%
\>[B]{}\Varid{second}\;\Varid{f}\;(\Varid{x},\Varid{y})\mathrel{=}\Varid{f}\;\Varid{y}\mathrel{\hstretch{0.7}{>\!\!>\!\!=}}\lambda \Varid{y'}\to \{(\Varid{x},\Varid{y'})\}~~.{}\<[E]%
\ColumnHook
\end{hscode}\resethooks
Our new wish is to construct an array counterpart of \ensuremath{\Varid{second}\;\Varid{perm}\mathbin{\cdot}\Varid{partl}\;\Varid{p}}.
Let the function be
\begin{hscode}\SaveRestoreHook
\column{B}{@{}>{\hspre}l<{\hspost}@{}}%
\column{5}{@{}>{\hspre}l<{\hspost}@{}}%
\column{E}{@{}>{\hspre}l<{\hspost}@{}}%
\>[B]{}\Varid{ipartl}\mathbin{::}(\Conid{MonadPlus}\;\Varid{m},\Conid{MonadArr}\;\Conid{Elm}\;\Varid{m})\Rightarrow {}\<[E]%
\\
\>[B]{}\hsindent{5}{}\<[5]%
\>[5]{}\Conid{Elm}\to \Conid{Int}\to (\Conid{Nat}\times\Conid{Nat}\times\Conid{Nat})\to \Varid{m}\;(\Conid{Nat}\times\Conid{Nat})~~.{}\<[E]%
\ColumnHook
\end{hscode}\resethooks
The intention is that in a call \ensuremath{\Varid{ipartl}\;\Varid{p}\;\Varid{i}\;(\Varid{ny},\Varid{nz},\Varid{nx})}, \ensuremath{\Varid{p}} is the pivot, \ensuremath{\Varid{i}} the index where \ensuremath{\Varid{ys}\mathbin{+\!\!\!+}\Varid{zs}\mathbin{+\!\!\!+}\bm{\Varid{xs}}} is stored in the array, and \ensuremath{\Varid{ny}}, \ensuremath{\Varid{nz}}, \ensuremath{\Varid{nx}} respectively the lengths of \ensuremath{\Varid{ys}}, \ensuremath{\Varid{zs}}, and \ensuremath{\Varid{xs}}. A specification of \ensuremath{\Varid{ipartl}} is:
\begin{hscode}\SaveRestoreHook
\column{B}{@{}>{\hspre}l<{\hspost}@{}}%
\column{4}{@{}>{\hspre}l<{\hspost}@{}}%
\column{E}{@{}>{\hspre}l<{\hspost}@{}}%
\>[B]{}\Varid{writeList}\;\Varid{i}\;(\Varid{ys}\mathbin{+\!\!\!+}\Varid{zs}\mathbin{+\!\!\!+}\bm{\Varid{xs}})\mathbin{\hstretch{0.7}{>\!\!>}}\Varid{ipartl}\;\Varid{p}\;\Varid{i}\;(\#\!\!\;\Varid{ys},\#\!\!\;\Varid{zs},\#\!\!\;\bm{\Varid{xs}})\mathrel{\subseteq}{}\<[E]%
\\
\>[B]{}\hsindent{4}{}\<[4]%
\>[4]{}\Varid{second}\;\Varid{perm}\;(\Varid{partl}\;\Varid{p}\;(\Varid{ys},\Varid{zs},\bm{\Varid{xs}}))\mathrel{\hstretch{0.7}{>\!\!>\!\!=}}\Varid{write2L}\;\Varid{i}~~.{}\<[E]%
\ColumnHook
\end{hscode}\resethooks
That is, under assumption that \ensuremath{\Varid{ys}\mathbin{+\!\!\!+}\Varid{zs}\mathbin{+\!\!\!+}\bm{\Varid{xs}}} is stored in the array starting from index \ensuremath{\Varid{i}} (initialised by \ensuremath{\Varid{writeList}}), \ensuremath{\Varid{ipartl}} computes \ensuremath{\Varid{partl}\;\Varid{p}\;(\Varid{ys},\Varid{zs},\bm{\Varid{xs}})}, possibly permuting the second partition.
The resulting two partitions are still stored in the array starting from \ensuremath{\Varid{i}}, and their lengths are returned.

\subsubsection{Derivation}
We start with fusing \ensuremath{\Varid{second}\;\Varid{perm}} into \ensuremath{\Varid{partl}},
that is, to construct \ensuremath{\Varid{partl'}\;\Varid{p}\mathrel{\dot{\subseteq}}\Varid{second}\;\Varid{perm}\mathbin{\cdot}\Varid{partl}\;\Varid{p}}.%
\footnote{We will discover a stronger specification \ensuremath{\Varid{partl'}\;\Varid{p}\mathrel{\dot{\subseteq}}\Varid{snd3}\;\Varid{perm}\mathbin{\backslash}(\Varid{second}\;\Varid{perm}\mathbin{\cdot}\Varid{partl}\;\Varid{p})},
where \ensuremath{\Varid{snd3}\;\Varid{f}\;(\Varid{x},\Varid{y},\Varid{z})\mathrel{=}\Varid{f}\;\Varid{y}\mathrel{\hstretch{0.7}{>\!\!>\!\!=}}\lambda \Varid{y'}\to \{(\Varid{x},\Varid{y'},\Varid{z})\}}.
We omit the details.
}
If we discover an inductive definition of \ensuremath{\Varid{partl'}}, it can then be used to construct an inductive definition of \ensuremath{\Varid{ipartl}}.
With some routine calculation we get:
\begin{hscode}\SaveRestoreHook
\column{B}{@{}>{\hspre}l<{\hspost}@{}}%
\column{3}{@{}>{\hspre}l<{\hspost}@{}}%
\column{15}{@{}>{\hspre}l<{\hspost}@{}}%
\column{21}{@{}>{\hspre}l<{\hspost}@{}}%
\column{28}{@{}>{\hspre}l<{\hspost}@{}}%
\column{E}{@{}>{\hspre}l<{\hspost}@{}}%
\>[B]{}\Varid{partl'}\mathbin{::}\Conid{MonadPlus}\;\Varid{m}\Rightarrow \Conid{Elm}\to (\Conid{List}\;\Conid{Elm})^3\to \Varid{m}\;(\Conid{List}\;\Conid{Elm}\times\Conid{List}\;\Conid{Elm}){}\<[E]%
\\
\>[B]{}\Varid{partl'}\;\Varid{p}\;(\Varid{ys},\Varid{zs},[\mskip1.5mu \mskip1.5mu]){}\<[28]%
\>[28]{}\mathrel{=}\{(\Varid{ys},\Varid{zs})\}{}\<[E]%
\\
\>[B]{}\Varid{partl'}\;\Varid{p}\;(\Varid{ys},\Varid{zs},\bm{\Varid{x}}\mathbin{:}\bm{\Varid{xs}}){}\<[28]%
\>[28]{}\mathrel{=}{}\<[E]%
\\
\>[B]{}\hsindent{3}{}\<[3]%
\>[3]{}\mathbf{if}\;\bm{\Varid{x}}\leq \Varid{p}\;{}\<[15]%
\>[15]{}\mathbf{then}\;{}\<[21]%
\>[21]{}\Varid{perm}\;\Varid{zs}\mathrel{\hstretch{0.7}{>\!\!>\!\!=}}\lambda \Varid{zs'}\to \Varid{partl'}\;\Varid{p}\;(\Varid{ys}\mathbin{+\!\!\!+}[\mskip1.5mu \bm{\Varid{x}}\mskip1.5mu],\Varid{zs'},\bm{\Varid{xs}}){}\<[E]%
\\
\>[15]{}\mathbf{else}\;{}\<[21]%
\>[21]{}\Varid{perm}\;(\Varid{zs}\mathbin{+\!\!\!+}[\mskip1.5mu \bm{\Varid{x}}\mskip1.5mu])\mathrel{\hstretch{0.7}{>\!\!>\!\!=}}\lambda \Varid{zs'}\to \Varid{partl'}\;\Varid{p}\;(\Varid{ys},\Varid{zs'},\bm{\Varid{xs}})~~.{}\<[E]%
\ColumnHook
\end{hscode}\resethooks
For an intuitive explanation, rather than permuting the second list \ensuremath{\Varid{zs}} after computing \ensuremath{\Varid{partl}}, we can also permute \ensuremath{\Varid{zs}} in \ensuremath{\Varid{partl'}} before every recursive call.

The specification of \ensuremath{\Varid{ipartl}} now becomes
\begin{equation}
\begin{split}
&\ensuremath{\Varid{writeList}\;\Varid{i}\;(\Varid{ys}\mathbin{+\!\!\!+}\Varid{zs}\mathbin{+\!\!\!+}\bm{\Varid{xs}})\mathbin{\hstretch{0.7}{>\!\!>}}\Varid{ipartl}\;\Varid{p}\;\Varid{i}\;(\#\!\!\;\Varid{ys},\#\!\!\;\Varid{zs},\#\!\!\;\bm{\Varid{xs}})\mathrel{\subseteq}} \\
&\qquad   \ensuremath{\Varid{partl'}\;\Varid{p}\;(\Varid{ys},\Varid{zs},\bm{\Varid{xs}})\mathrel{\hstretch{0.7}{>\!\!>\!\!=}}\Varid{write2L}\;\Varid{i}} \mbox{~~.}
   \label{eq:ipartl-spec}
\end{split}
\end{equation}
To calculate \ensuremath{\Varid{ipartl}}, we start with the right-hand side of \ensuremath{(\mathrel{\subseteq})},
since it contains more information to work with.
We try to push \ensuremath{\Varid{write2L}} leftwards until the expression has the form \ensuremath{\Varid{writeList}\;\Varid{i}\;(\Varid{ys}\mathbin{+\!\!\!+}\Varid{zs}\mathbin{+\!\!\!+}\Varid{xs})\mathbin{\hstretch{0.7}{>\!\!>}}\mathbin{...}}, thereby constructing \ensuremath{\Varid{ipartl}}.
This is similar to that, in imperative program calculation, we {\em work backwards from the postcondition} to construct a program that works under the given precondition~\cite{Dijkstra:76:Discipline}.

We intend to construct \ensuremath{\Varid{ipartl}} by induction on \ensuremath{\Varid{xs}}.
For \ensuremath{\Varid{xs}\mathbin{:=}[\mskip1.5mu \mskip1.5mu]}, we get \ensuremath{\Varid{ipartl}\;\Varid{p}\;\Varid{i}} \ensuremath{(\Varid{ny},\Varid{nz},\mathrm{0})\mathrel{=}\{(\Varid{ny},\Varid{nz})\}}.
For the case \ensuremath{\Varid{x}\mathbin{:}\Varid{xs}}, assume that the specification is met for \ensuremath{\Varid{xs}}.
Just for making the calculation shorter, we refactor \ensuremath{\Varid{partl'}}, lifting the recursive calls and turning the main body into an auxiliary function:
\begin{hscode}\SaveRestoreHook
\column{B}{@{}>{\hspre}l<{\hspost}@{}}%
\column{3}{@{}>{\hspre}l<{\hspost}@{}}%
\column{10}{@{}>{\hspre}l<{\hspost}@{}}%
\column{22}{@{}>{\hspre}l<{\hspost}@{}}%
\column{28}{@{}>{\hspre}l<{\hspost}@{}}%
\column{E}{@{}>{\hspre}l<{\hspost}@{}}%
\>[B]{}\Varid{partl'}\;\Varid{p}\;(\Varid{ys},\Varid{zs},\bm{\Varid{x}}\mathbin{:}\bm{\Varid{xs}})\mathbin{=}\Varid{dispatch}\;\bm{\Varid{x}}\;\Varid{p}\;(\Varid{ys},\Varid{zs},\bm{\Varid{xs}})\mathrel{\hstretch{0.7}{>\!\!>\!\!=}}\Varid{partl'}\;\Varid{p}~~,{}\<[E]%
\\
\>[B]{}\hsindent{3}{}\<[3]%
\>[3]{}\mathbf{where}\;\Varid{dispatch}\;\bm{\Varid{x}}\;\Varid{p}\;(\Varid{ys},\Varid{zs},\bm{\Varid{xs}})\mathrel{=}{}\<[E]%
\\
\>[3]{}\hsindent{7}{}\<[10]%
\>[10]{}\mathbf{if}\;\bm{\Varid{x}}\leq \Varid{p}\;{}\<[22]%
\>[22]{}\mathbf{then}\;{}\<[28]%
\>[28]{}\Varid{perm}\;\Varid{zs}\mathrel{\hstretch{0.7}{>\!\!>\!\!=}}\lambda \Varid{zs'}\to \{(\Varid{ys}\mathbin{+\!\!\!+}[\mskip1.5mu \bm{\Varid{x}}\mskip1.5mu],\Varid{zs'},\bm{\Varid{xs}})\}{}\<[E]%
\\
\>[22]{}\mathbf{else}\;{}\<[28]%
\>[28]{}\Varid{perm}\;(\Varid{zs}\mathbin{+\!\!\!+}[\mskip1.5mu \bm{\Varid{x}}\mskip1.5mu])\mathrel{\hstretch{0.7}{>\!\!>\!\!=}}\lambda \Varid{zs'}\to \{(\Varid{ys},\Varid{zs'},\bm{\Varid{xs}})\}~~.{}\<[E]%
\ColumnHook
\end{hscode}\resethooks
We calculate:
\begin{hscode}\SaveRestoreHook
\column{B}{@{}>{\hspre}l<{\hspost}@{}}%
\column{4}{@{}>{\hspre}l<{\hspost}@{}}%
\column{5}{@{}>{\hspre}l<{\hspost}@{}}%
\column{12}{@{}>{\hspre}l<{\hspost}@{}}%
\column{16}{@{}>{\hspre}l<{\hspost}@{}}%
\column{22}{@{}>{\hspre}l<{\hspost}@{}}%
\column{E}{@{}>{\hspre}l<{\hspost}@{}}%
\>[5]{}\Varid{partl'}\;\Varid{p}\;(\Varid{ys},\Varid{zs},\bm{\Varid{x}}\mathbin{:}\bm{\Varid{xs}})\mathrel{\hstretch{0.7}{>\!\!>\!\!=}}\Varid{write2L}\;\Varid{i}{}\<[E]%
\\
\>[B]{}\mathbin{=}{}\<[12]%
\>[12]{}\mbox{\commentbegin  definition of \ensuremath{\Varid{partl'}}  \commentend}{}\<[E]%
\\
\>[B]{}\hsindent{5}{}\<[5]%
\>[5]{}(\Varid{dispatch}\;\bm{\Varid{x}}\;\Varid{p}\;(\Varid{ys},\Varid{zs},\bm{\Varid{xs}})\mathrel{\hstretch{0.7}{>\!\!>\!\!=}}\Varid{partl'}\;\Varid{p})\mathrel{\hstretch{0.7}{>\!\!>\!\!=}}\Varid{write2L}\;\Varid{i}{}\<[E]%
\\
\>[B]{}\mathrel{\supseteq}{}\<[12]%
\>[12]{}\mbox{\commentbegin  monad laws, inductive assumption  \commentend}{}\<[E]%
\\
\>[B]{}\hsindent{5}{}\<[5]%
\>[5]{}(\Varid{dispatch}\;\bm{\Varid{x}}\;\Varid{p}\;(\Varid{ys},\Varid{zs},\bm{\Varid{xs}})\mathrel{\hstretch{0.7}{>\!\!>\!\!=}}\Varid{write3L}\;\Varid{i})\mathrel{\hstretch{0.7}{>\!\!>\!\!=}}\Varid{ipartl}\;\Varid{p}\;\Varid{i}{}\<[E]%
\\
\>[B]{}\mathbin{=}{}\<[12]%
\>[12]{}\mbox{\commentbegin  by \eqref{eq:writeList-++}, monad laws  \commentend}{}\<[E]%
\\
\>[B]{}\hsindent{5}{}\<[5]%
\>[5]{}\Varid{dispatch}\;\bm{\Varid{x}}\;\Varid{p}\;(\Varid{ys},\Varid{zs},\bm{\Varid{xs}})\mathrel{\hstretch{0.7}{>\!\!>\!\!=}}\lambda (\Varid{ys'},\Varid{zs'},\bm{\Varid{xs}})\to {}\<[E]%
\\
\>[B]{}\hsindent{5}{}\<[5]%
\>[5]{}\Varid{writeList}\;\Varid{i}\;(\Varid{ys'}\mathbin{+\!\!\!+}\Varid{zs'})\mathbin{\hstretch{0.7}{>\!\!>}}\Varid{writeList}\;(\Varid{i}\mathbin{+}\#\!\!\;(\Varid{ys'}\mathbin{+\!\!\!+}\Varid{zs'}))\;\bm{\Varid{xs}}\mathbin{\hstretch{0.7}{>\!\!>}}{}\<[E]%
\\
\>[B]{}\hsindent{5}{}\<[5]%
\>[5]{}\Varid{ipartl}\;\Varid{p}\;\Varid{i}\;(\#\!\!\;\Varid{ys'},\#\!\!\;\Varid{zs'},\#\!\!\;\bm{\Varid{xs}}){}\<[E]%
\\
\>[B]{}\mathbin{=}{}\<[12]%
\>[12]{}\mbox{\commentbegin  \ensuremath{\Varid{perm}} preserves length, commutativity  \commentend}{}\<[E]%
\\
\>[B]{}\hsindent{5}{}\<[5]%
\>[5]{}\Varid{writeList}\;(\Varid{i}\mathbin{+}\#\!\!\;\Varid{ys}\mathbin{+}\#\!\!\;\Varid{zs}\mathbin{+}\mathrm{1})\;\bm{\Varid{xs}}\mathbin{\hstretch{0.7}{>\!\!>}}{}\<[E]%
\\
\>[B]{}\hsindent{5}{}\<[5]%
\>[5]{}\Varid{dispatch}\;\bm{\Varid{x}}\;\Varid{p}\;(\Varid{ys},\Varid{zs},\bm{\Varid{xs}})\mathrel{\hstretch{0.7}{>\!\!>\!\!=}}\lambda (\Varid{ys'},\Varid{zs'},\bm{\Varid{xs}})\to {}\<[E]%
\\
\>[B]{}\hsindent{5}{}\<[5]%
\>[5]{}\Varid{writeList}\;\Varid{i}\;(\Varid{ys'}\mathbin{+\!\!\!+}\Varid{zs'})\mathbin{\hstretch{0.7}{>\!\!>}}{}\<[E]%
\\
\>[B]{}\hsindent{5}{}\<[5]%
\>[5]{}\Varid{ipartl}\;\Varid{p}\;\Varid{i}\;(\#\!\!\;\Varid{ys'},\#\!\!\;\Varid{zs'},\#\!\!\;\bm{\Varid{xs}}){}\<[E]%
\\
\>[B]{}\mathbin{=}{}\<[12]%
\>[12]{}\mbox{\commentbegin  definition of \ensuremath{\Varid{dispatch}}, function calls distribute into \ensuremath{\mathbf{if}}  \commentend}{}\<[E]%
\\
\>[B]{}\hsindent{4}{}\<[4]%
\>[4]{}\Varid{writeList}\;(\Varid{i}\mathbin{+}\#\!\!\;\Varid{ys}\mathbin{+}\#\!\!\;\Varid{zs}\mathbin{+}\mathrm{1})\;\bm{\Varid{xs}}\mathbin{\hstretch{0.7}{>\!\!>}}{}\<[E]%
\\
\>[B]{}\hsindent{4}{}\<[4]%
\>[4]{}\mathbf{if}\;\bm{\Varid{x}}\leq \Varid{p}\;{}\<[16]%
\>[16]{}\mathbf{then}\;{}\<[22]%
\>[22]{}\Varid{perm}\;\Varid{zs}\mathrel{\hstretch{0.7}{>\!\!>\!\!=}}\lambda \Varid{zs'}\to \Varid{writeList}\;\Varid{i}\;(\Varid{ys}\mathbin{+\!\!\!+}[\mskip1.5mu \bm{\Varid{x}}\mskip1.5mu]\mathbin{+\!\!\!+}\Varid{zs'})\mathbin{\hstretch{0.7}{>\!\!>}}{}\<[E]%
\\
\>[22]{}\Varid{ipartl}\;\Varid{p}\;\Varid{i}\;(\#\!\!\;\Varid{ys}\mathbin{+}\mathrm{1},\#\!\!\;\Varid{zs'},\#\!\!\;\bm{\Varid{xs}}){}\<[E]%
\\
\>[16]{}\mathbf{else}\;{}\<[22]%
\>[22]{}\Varid{perm}\;(\Varid{zs}\mathbin{+\!\!\!+}[\mskip1.5mu \bm{\Varid{x}}\mskip1.5mu])\mathrel{\hstretch{0.7}{>\!\!>\!\!=}}\lambda \Varid{zs'}\to \Varid{writeList}\;\Varid{i}\;(\Varid{ys}\mathbin{+\!\!\!+}\Varid{zs'})\mathbin{\hstretch{0.7}{>\!\!>}}{}\<[E]%
\\
\>[22]{}\Varid{ipartl}\;\Varid{p}\;\Varid{i}\;(\#\!\!\;\Varid{ys},\#\!\!\;\Varid{zs'},\#\!\!\;\bm{\Varid{xs}})~~.{}\<[E]%
\ColumnHook
\end{hscode}\resethooks

We pause here to see what has happened: we have constructed a precondition \ensuremath{\Varid{writeList}\;(\Varid{i}\mathbin{+}\#\!\!\;\Varid{ys}\mathbin{+}\#\!\!\;\Varid{zs}\mathbin{+}\mathrm{1})\;\bm{\Varid{xs}}}, which is part of the desired precondition: \ensuremath{\Varid{writeList}\;\Varid{i}\;(\Varid{ys}\mathbin{+\!\!\!+}\Varid{zs}\mathbin{+\!\!\!+}(\bm{\Varid{x}}\mathbin{:}\bm{\Varid{xs}}))}.
To recover the latter precondition, we will try to turn both branches of \ensuremath{\mathbf{if}} into the form \ensuremath{\Varid{writeList}\;\Varid{i}\;(\Varid{ys}\mathbin{+\!\!\!+}\Varid{zs}\mathbin{+\!\!\!+}[\mskip1.5mu \bm{\Varid{x}}\mskip1.5mu])\mathrel{\hstretch{0.7}{>\!\!>\!\!=}}\mathbin{...}}. That is, we try to construct, in both branches, some code that executes under the precondition \ensuremath{\Varid{writeList}\;\Varid{i}\;(\Varid{ys}\mathbin{+\!\!\!+}\Varid{zs}\mathbin{+\!\!\!+}[\mskip1.5mu \bm{\Varid{x}}\mskip1.5mu])} --- that the code generates the correct result is guaranteed by the refinement relation.

It is easier for the second branch, where we can simply refine \ensuremath{\Varid{perm}} to \ensuremath{\{\cdot \}}:
\begin{hscode}\SaveRestoreHook
\column{B}{@{}>{\hspre}l<{\hspost}@{}}%
\column{8}{@{}>{\hspre}l<{\hspost}@{}}%
\column{11}{@{}>{\hspre}l<{\hspost}@{}}%
\column{E}{@{}>{\hspre}l<{\hspost}@{}}%
\>[8]{}\Varid{perm}\;(\Varid{zs}\mathbin{+\!\!\!+}[\mskip1.5mu \bm{\Varid{x}}\mskip1.5mu])\mathrel{\hstretch{0.7}{>\!\!>\!\!=}}\lambda \Varid{zs'}\to \Varid{writeList}\;\Varid{i}\;(\Varid{ys}\mathbin{+\!\!\!+}\Varid{zs'})\mathbin{\hstretch{0.7}{>\!\!>}}{}\<[E]%
\\
\>[8]{}\Varid{ipartl}\;\Varid{p}\;\Varid{i}\;(\#\!\!\;\Varid{ys},\#\!\!\;\Varid{zs'},\#\!\!\;\bm{\Varid{xs}}){}\<[E]%
\\
\>[B]{}\mathrel{\supseteq}{}\<[11]%
\>[11]{}\mbox{\commentbegin  since \ensuremath{\{\Varid{xs}\}\mathrel{\subseteq}\Varid{perm}\;\Varid{xs}}  \commentend}{}\<[E]%
\\
\>[B]{}\hsindent{8}{}\<[8]%
\>[8]{}\Varid{writeList}\;\Varid{i}\;(\Varid{ys}\mathbin{+\!\!\!+}\Varid{zs}\mathbin{+\!\!\!+}[\mskip1.5mu \bm{\Varid{x}}\mskip1.5mu])\mathbin{\hstretch{0.7}{>\!\!>}}\Varid{ipartl}\;\Varid{p}\;\Varid{i}\;(\#\!\!\;\Varid{ys},\#\!\!\;\Varid{zs}\mathbin{+}\mathrm{1},\#\!\!\;\bm{\Varid{xs}})~~.{}\<[E]%
\ColumnHook
\end{hscode}\resethooks

For the first branch, we focus on its first line:
\begin{hscode}\SaveRestoreHook
\column{B}{@{}>{\hspre}l<{\hspost}@{}}%
\column{4}{@{}>{\hspre}l<{\hspost}@{}}%
\column{11}{@{}>{\hspre}l<{\hspost}@{}}%
\column{E}{@{}>{\hspre}l<{\hspost}@{}}%
\>[4]{}\Varid{perm}\;\Varid{zs}\mathrel{\hstretch{0.7}{>\!\!>\!\!=}}\lambda \Varid{zs'}\to \Varid{writeList}\;\Varid{i}\;(\Varid{ys}\mathbin{+\!\!\!+}[\mskip1.5mu \bm{\Varid{x}}\mskip1.5mu]\mathbin{+\!\!\!+}\Varid{zs'}){}\<[E]%
\\
\>[B]{}\mathbin{=}{}\<[11]%
\>[11]{}\mbox{\commentbegin  by \eqref{eq:writeList-++}, commutativity  \commentend}{}\<[E]%
\\
\>[B]{}\hsindent{4}{}\<[4]%
\>[4]{}\Varid{writeList}\;\Varid{i}\;\Varid{ys}\mathbin{\hstretch{0.7}{>\!\!>}}\Varid{perm}\;\Varid{zs}\mathrel{\hstretch{0.7}{>\!\!>\!\!=}}\lambda \Varid{zs'}\to \Varid{writeList}\;(\Varid{i}\mathbin{+}\#\!\!\;\Varid{ys})\;([\mskip1.5mu \bm{\Varid{x}}\mskip1.5mu]\mathbin{+\!\!\!+}\Varid{zs'}){}\<[E]%
\\
\>[B]{}\mathrel{\supseteq}{}\<[11]%
\>[11]{}\mbox{\commentbegin  introduce \ensuremath{\Varid{swap}}, see below  \commentend}{}\<[E]%
\\
\>[B]{}\hsindent{4}{}\<[4]%
\>[4]{}\Varid{writeList}\;\Varid{i}\;\Varid{ys}\mathbin{\hstretch{0.7}{>\!\!>}}\Varid{writeList}\;(\Varid{i}\mathbin{+}\#\!\!\;\Varid{ys})\;(\Varid{zs}\mathbin{+\!\!\!+}[\mskip1.5mu \bm{\Varid{x}}\mskip1.5mu])\mathbin{\hstretch{0.7}{>\!\!>}}{}\<[E]%
\\
\>[B]{}\hsindent{4}{}\<[4]%
\>[4]{}\Varid{swap}\;(\Varid{i}\mathbin{+}\#\!\!\;\Varid{ys})\;(\Varid{i}\mathbin{+}\#\!\!\;\Varid{ys}\mathbin{+}\#\!\!\;\Varid{zs}){}\<[E]%
\\
\>[B]{}\mathbin{=}{}\<[11]%
\>[11]{}\mbox{\commentbegin  by \eqref{eq:writeList-++}  \commentend}{}\<[E]%
\\
\>[B]{}\hsindent{4}{}\<[4]%
\>[4]{}\Varid{writeList}\;\Varid{i}\;(\Varid{ys}\mathbin{+\!\!\!+}\Varid{zs}\mathbin{+\!\!\!+}[\mskip1.5mu \bm{\Varid{x}}\mskip1.5mu])\mathbin{\hstretch{0.7}{>\!\!>}}\Varid{swap}\;(\Varid{i}\mathbin{+}\#\!\!\;\Varid{ys})\;(\Varid{i}\mathbin{+}\#\!\!\;\Varid{ys}\mathbin{+}\#\!\!\;\Varid{zs})~~.{}\<[E]%
\ColumnHook
\end{hscode}\resethooks
Here we explain the last two steps.
Operationally speaking, given an array containing \ensuremath{\Varid{ys}\mathbin{+\!\!\!+}\Varid{zs}\mathbin{+\!\!\!+}[\mskip1.5mu \bm{\Varid{x}}\mskip1.5mu]} (the precondition we wanted, initialized by the \ensuremath{\Varid{writeList}} in the last line), how do we mutate it to \ensuremath{\Varid{ys}\mathbin{+\!\!\!+}[\mskip1.5mu \bm{\Varid{x}}\mskip1.5mu]\mathbin{+\!\!\!+}\Varid{zs'}} (postcondition specified by the \ensuremath{\Varid{writeList}} in the first line), where \ensuremath{\Varid{zs'}} is a permutation of \ensuremath{\Varid{zs}}? We may do so by swapping \ensuremath{\bm{\Varid{x}}} with the leftmost element of \ensuremath{\Varid{zs}}, which is what we did in the second step. Formally, we used the property:
\begin{equation}
\begin{split}
&\ensuremath{\Varid{perm}\;\Varid{zs}\mathrel{\hstretch{0.7}{>\!\!>\!\!=}}\lambda \Varid{zs'}\to \Varid{writeList}\;\Varid{i}\;([\mskip1.5mu \bm{\Varid{x}}\mskip1.5mu]\mathbin{+\!\!\!+}\Varid{zs'})\mathrel{\supseteq}}\\
&\qquad\ensuremath{\Varid{writeList}\;\Varid{i}\;(\Varid{zs}\mathbin{+\!\!\!+}[\mskip1.5mu \bm{\Varid{x}}\mskip1.5mu])\mathbin{\hstretch{0.7}{>\!\!>}}\Varid{swap}\;\Varid{i}\;(\Varid{i}\mathbin{+}\#\!\!\;\Varid{zs})~~.}
\end{split} \label{eq:perm-write-swap}
\end{equation}

Now that both branches are refined to code with precondition \ensuremath{\Varid{writeList}\;\Varid{i}\;(\Varid{ys}\mathbin{+\!\!\!+}\Varid{zs}\mathbin{+\!\!\!+}[\mskip1.5mu \bm{\Varid{x}}\mskip1.5mu])}, we go back to the main derivation:
\begin{hscode}\SaveRestoreHook
\column{B}{@{}>{\hspre}l<{\hspost}@{}}%
\column{3}{@{}>{\hspre}l<{\hspost}@{}}%
\column{8}{@{}>{\hspre}l<{\hspost}@{}}%
\column{15}{@{}>{\hspre}l<{\hspost}@{}}%
\column{21}{@{}>{\hspre}l<{\hspost}@{}}%
\column{E}{@{}>{\hspre}l<{\hspost}@{}}%
\>[3]{}\Varid{writeList}\;(\Varid{i}\mathbin{+}\#\!\!\;\Varid{ys}\mathbin{+}\#\!\!\;\Varid{zs}\mathbin{+}\mathrm{1})\;\bm{\Varid{xs}}\mathbin{\hstretch{0.7}{>\!\!>}}{}\<[E]%
\\
\>[3]{}\mathbf{if}\;\bm{\Varid{x}}\leq \Varid{p}\;{}\<[15]%
\>[15]{}\mathbf{then}\;{}\<[21]%
\>[21]{}\Varid{writeList}\;\Varid{i}\;(\Varid{ys}\mathbin{+\!\!\!+}\Varid{zs}\mathbin{+\!\!\!+}[\mskip1.5mu \bm{\Varid{x}}\mskip1.5mu])\mathbin{\hstretch{0.7}{>\!\!>}}{}\<[E]%
\\
\>[21]{}\Varid{swap}\;(\Varid{i}\mathbin{+}\#\!\!\;\Varid{ys})\;(\Varid{i}\mathbin{+}\#\!\!\;\Varid{ys}\mathbin{+}\#\!\!\;\Varid{zs})\mathbin{\hstretch{0.7}{>\!\!>}}{}\<[E]%
\\
\>[21]{}\Varid{ipartl}\;\Varid{p}\;\Varid{i}\;(\#\!\!\;\Varid{ys}\mathbin{+}\mathrm{1},\#\!\!\;\Varid{zs},\#\!\!\;\bm{\Varid{xs}}){}\<[E]%
\\
\>[15]{}\mathbf{else}\;{}\<[21]%
\>[21]{}\Varid{writeList}\;\Varid{i}\;(\Varid{ys}\mathbin{+\!\!\!+}\Varid{zs}\mathbin{+\!\!\!+}[\mskip1.5mu \bm{\Varid{x}}\mskip1.5mu])\mathbin{\hstretch{0.7}{>\!\!>}}{}\<[E]%
\\
\>[21]{}\Varid{ipartl}\;\Varid{p}\;\Varid{i}\;(\#\!\!\;\Varid{ys},\#\!\!\;\Varid{zs}\mathbin{+}\mathrm{1},\#\!\!\;\bm{\Varid{xs}}){}\<[E]%
\\
\>[B]{}\mathbin{=}{}\<[8]%
\>[8]{}\mbox{\commentbegin  distributivity of \ensuremath{\mathbf{if}}, \eqref{eq:writeList-++}  \commentend}{}\<[E]%
\\
\>[B]{}\hsindent{3}{}\<[3]%
\>[3]{}\Varid{writeList}\;\Varid{i}\;(\Varid{ys}\mathbin{+\!\!\!+}\Varid{zs}\mathbin{+\!\!\!+}(\bm{\Varid{x}}\mathbin{:}\bm{\Varid{xs}}))\mathbin{\hstretch{0.7}{>\!\!>}}{}\<[E]%
\\
\>[B]{}\hsindent{3}{}\<[3]%
\>[3]{}\mathbf{if}\;\bm{\Varid{x}}\leq \Varid{p}\;{}\<[15]%
\>[15]{}\mathbf{then}\;{}\<[21]%
\>[21]{}\Varid{swap}\;(\Varid{i}\mathbin{+}\#\!\!\;\Varid{ys})\;(\Varid{i}\mathbin{+}\#\!\!\;\Varid{ys}\mathbin{+}\#\!\!\;\Varid{zs})\mathbin{\hstretch{0.7}{>\!\!>}}{}\<[E]%
\\
\>[21]{}\Varid{ipartl}\;\Varid{p}\;\Varid{i}\;(\#\!\!\;\Varid{ys}\mathbin{+}\mathrm{1},\#\!\!\;\Varid{zs},\#\!\!\;\bm{\Varid{xs}}){}\<[E]%
\\
\>[15]{}\mathbf{else}\;{}\<[21]%
\>[21]{}\Varid{ipartl}\;\Varid{p}\;\Varid{i}\;(\#\!\!\;\Varid{ys},\#\!\!\;\Varid{zs}\mathbin{+}\mathrm{1},\#\!\!\;\bm{\Varid{xs}}){}\<[E]%
\\
\>[B]{}\mathbin{=}{}\<[8]%
\>[8]{}\mbox{\commentbegin  {\bf write-read} and definition of \ensuremath{\Varid{writeList}}  \commentend}{}\<[E]%
\\
\>[B]{}\hsindent{3}{}\<[3]%
\>[3]{}\Varid{writeList}\;\Varid{i}\;(\Varid{ys}\mathbin{+\!\!\!+}\Varid{zs}\mathbin{+\!\!\!+}(\bm{\Varid{x}}\mathbin{:}\bm{\Varid{xs}}))\mathbin{\hstretch{0.7}{>\!\!>}}{}\<[E]%
\\
\>[B]{}\hsindent{3}{}\<[3]%
\>[3]{}\Varid{read}\;(\Varid{i}\mathbin{+}\#\!\!\;\Varid{ys}\mathbin{+}\#\!\!\;\Varid{zs})\mathrel{\hstretch{0.7}{>\!\!>\!\!=}}\lambda \bm{\Varid{x}}\to {}\<[E]%
\\
\>[B]{}\hsindent{3}{}\<[3]%
\>[3]{}\mathbf{if}\;\bm{\Varid{x}}\leq \Varid{p}\;{}\<[15]%
\>[15]{}\mathbf{then}\;{}\<[21]%
\>[21]{}\Varid{swap}\;(\Varid{i}\mathbin{+}\#\!\!\;\Varid{ys})\;(\Varid{i}\mathbin{+}\#\!\!\;\Varid{ys}\mathbin{+}\#\!\!\;\Varid{zs})\mathbin{\hstretch{0.7}{>\!\!>}}{}\<[E]%
\\
\>[21]{}\Varid{ipartl}\;\Varid{p}\;\Varid{i}\;(\#\!\!\;\Varid{ys}\mathbin{+}\mathrm{1},\#\!\!\;\Varid{zs},\#\!\!\;\bm{\Varid{xs}}){}\<[E]%
\\
\>[15]{}\mathbf{else}\;{}\<[21]%
\>[21]{}\Varid{ipartl}\;\Varid{p}\;\Varid{i}\;(\#\!\!\;\Varid{ys},\#\!\!\;\Varid{zs}\mathbin{+}\mathrm{1},\#\!\!\;\bm{\Varid{xs}})~~.{}\<[E]%
\ColumnHook
\end{hscode}\resethooks
We have thus established the precondition \ensuremath{\Varid{writeList}\;\Varid{i}\;(\Varid{ys}\mathbin{+\!\!\!+}\Varid{zs}\mathbin{+\!\!\!+}(\bm{\Varid{x}}\mathbin{:}\bm{\Varid{xs}}))}.
In summary, we have derived:
\begin{hscode}\SaveRestoreHook
\column{B}{@{}>{\hspre}l<{\hspost}@{}}%
\column{3}{@{}>{\hspre}l<{\hspost}@{}}%
\column{12}{@{}>{\hspre}l<{\hspost}@{}}%
\column{18}{@{}>{\hspre}l<{\hspost}@{}}%
\column{27}{@{}>{\hspre}l<{\hspost}@{}}%
\column{E}{@{}>{\hspre}l<{\hspost}@{}}%
\>[B]{}\Varid{ipartl}\mathbin{::}\Conid{MonadArr}\;\Conid{Elm}\;\Varid{m}\Rightarrow \Conid{Elm}\to \Conid{Int}\to (\Conid{Int}\times\Conid{Int}\times\Conid{Int})\to \Varid{m}\;(\Conid{Int}\times\Conid{Int}){}\<[E]%
\\
\>[B]{}\Varid{ipartl}\;\Varid{p}\;\Varid{i}\;(\Varid{ny},\Varid{nz},\mathrm{0}){}\<[27]%
\>[27]{}\mathrel{=}\{(\Varid{ny},\Varid{nz})\}{}\<[E]%
\\
\>[B]{}\Varid{ipartl}\;\Varid{p}\;\Varid{i}\;(\Varid{ny},\Varid{nz},\mathrm{1}\mathbin{+}\Varid{k}){}\<[27]%
\>[27]{}\mathrel{=}{}\<[E]%
\\
\>[B]{}\hsindent{3}{}\<[3]%
\>[3]{}\Varid{read}\;(\Varid{i}\mathbin{+}\Varid{ny}\mathbin{+}\Varid{nz})\mathrel{\hstretch{0.7}{>\!\!>\!\!=}}\lambda \Varid{x}\to {}\<[E]%
\\
\>[B]{}\hsindent{3}{}\<[3]%
\>[3]{}\mathbf{if}\;\Varid{x}\leq \Varid{p}\;{}\<[12]%
\>[12]{}\mathbf{then}\;{}\<[18]%
\>[18]{}\Varid{swap}\;(\Varid{i}\mathbin{+}\Varid{ny})\;(\Varid{i}\mathbin{+}\Varid{ny}\mathbin{+}\Varid{nz})\mathbin{\hstretch{0.7}{>\!\!>}}\Varid{ipartl}\;\Varid{p}\;\Varid{i}\;(\Varid{ny}\mathbin{+}\mathrm{1},\Varid{nz},\Varid{k}){}\<[E]%
\\
\>[12]{}\mathbf{else}\;{}\<[18]%
\>[18]{}\Varid{ipartl}\;\Varid{p}\;\Varid{i}\;(\Varid{ny},\Varid{nz}\mathbin{+}\mathrm{1},\Varid{k})~~.{}\<[E]%
\ColumnHook
\end{hscode}\resethooks

\subsection{Sorting an Array}

Now that we have \ensuremath{\Varid{ipartl}} derived, the rest of the work is to install it into quicksort.
We intend to derive \ensuremath{\Varid{iqsort}\mathbin{::}\Conid{MonadArr}\;\Conid{Elm}\;\Varid{m}\Rightarrow \Conid{Int}\to \Conid{Nat}\to \Varid{m}\;()} such that \ensuremath{\Varid{isort}\;\Varid{i}\;\Varid{n}} sorts the \ensuremath{\Varid{n}} elements in the array starting from index \ensuremath{\Varid{i}}.
We can give it a formal specification:
\begin{align}
\ensuremath{\Varid{writeList}\;\Varid{i}\;\Varid{xs}\mathbin{\hstretch{0.7}{>\!\!>}}\Varid{iqsort}\;\Varid{i}\;(\#\!\!\;\Varid{xs})} ~\subseteq~
  \ensuremath{\Varid{slowsort}\;\Varid{xs}\mathrel{\hstretch{0.7}{>\!\!>\!\!=}}\Varid{writeList}\;\Varid{i}} \mbox{~~.}\label{eq:iqsort-spec}
\end{align}
That is, when \ensuremath{\Varid{iqsort}\;\Varid{i}} is run from a state initialised by \ensuremath{\Varid{writeList}\;\Varid{i}\;\Varid{xs}}, it should behave the same as \ensuremath{\Varid{slowsort}\;\Varid{xs}\mathrel{\hstretch{0.7}{>\!\!>\!\!=}}\Varid{writeList}\;\Varid{i}}.

The function \ensuremath{\Varid{iqsort}} can be constructed by induction on the length of the input list.
For the case \ensuremath{\Varid{xs}\mathbin{:=}\Varid{p}\mathbin{:}\Varid{xs}}, we start from the left-hand side
\ensuremath{\Varid{slowsort}\;(\Varid{p}\mathbin{:}\Varid{xs})\mathrel{\hstretch{0.7}{>\!\!>\!\!=}}\Varid{writeList}\;\Varid{i}} and attempt to transform it to
\ensuremath{\Varid{writeList}\;\Varid{i}\;(\Varid{p}\mathbin{:}\Varid{xs})\mathbin{\hstretch{0.7}{>\!\!>}}\mathbin{...}}, thereby construct \ensuremath{\Varid{iqsort}}.
We present only the hightlights of the derivation.
Firstly, \ensuremath{\Varid{slowsort}\;(\Varid{p}\mathbin{:}\Varid{xs})\mathrel{\hstretch{0.7}{>\!\!>\!\!=}}\Varid{writeList}\;\Varid{i}} can be transformed to:
\begin{hscode}\SaveRestoreHook
\column{B}{@{}>{\hspre}l<{\hspost}@{}}%
\column{E}{@{}>{\hspre}l<{\hspost}@{}}%
\>[B]{}\Varid{partl'}\;\Varid{p}\;([\mskip1.5mu \mskip1.5mu],[\mskip1.5mu \mskip1.5mu],\Varid{xs})\mathrel{\hstretch{0.7}{>\!\!>\!\!=}}\lambda (\Varid{ys},\Varid{zs})\to {}\<[E]%
\\
\>[B]{}\Varid{perm}\;\Varid{ys}\mathrel{\hstretch{0.7}{>\!\!>\!\!=}}\lambda \Varid{ys'}\to \Varid{writeList}\;\Varid{i}\;(\Varid{ys'}\mathbin{+\!\!\!+}[\mskip1.5mu \Varid{p}\mskip1.5mu]\mathbin{+\!\!\!+}\Varid{zs})\mathbin{\hstretch{0.7}{>\!\!>}}{}\<[E]%
\\
\>[B]{}\Varid{iqsort}\;\Varid{i}\;(\#\!\!\;\Varid{ys})\mathbin{\hstretch{0.7}{>\!\!>}}\Varid{iqsort}\;(\Varid{i}\mathbin{+}\#\!\!\;\Varid{ys}\mathbin{+}\mathrm{1})\;(\#\!\!\;\Varid{zs})~~.{}\<[E]%
\ColumnHook
\end{hscode}\resethooks
For that to work, we introduced two \ensuremath{\Varid{perm}} to permute both partitions generated by \ensuremath{\Varid{partition}}. We can do so because \ensuremath{\Varid{perm}\mathrel{\hstretch{0.7}{>\!\!=\!\!\!>}}\Varid{perm}\mathrel{=}\Varid{perm}} and thus \ensuremath{\Varid{perm}\mathrel{\hstretch{0.7}{>\!\!=\!\!\!>}}\Varid{slowsort}\mathrel{=}\Varid{slowsort}}. The term \ensuremath{\Varid{perm}\;\Varid{zs}} was combined with \ensuremath{\Varid{partition}\;\Varid{p}}, yielding \ensuremath{\Varid{partl'}\;\Varid{p}}, while \ensuremath{\Varid{perm}\;\Varid{ys}} will be needed later.
We also needed \eqref{eq:writeList-++} to split \ensuremath{\Varid{writeList}\;\Varid{i}\;(\Varid{ys'}\mathbin{+\!\!\!+}[\mskip1.5mu \Varid{x}\mskip1.5mu]\mathbin{+\!\!\!+}\Varid{zs'})} into two parts. Assuming that \eqref{eq:iqsort-spec} has been met for lists shorter than \ensuremath{\Varid{xs}}, two subexpressions are folded back to \ensuremath{\Varid{iqsort}}.


Now that we have introduced \ensuremath{\Varid{partl'}}, the next goal is to embed \ensuremath{\Varid{ipartl}}.
The status of the array before the two calls to \ensuremath{\Varid{iqsort}} is given by \ensuremath{\Varid{writeList}\;\Varid{i}\;(\Varid{ys'}\mathbin{+\!\!\!+}[\mskip1.5mu \Varid{p}\mskip1.5mu]\mathbin{+\!\!\!+}\Varid{zs})}. That is, \ensuremath{\Varid{ys'}\mathbin{+\!\!\!+}[\mskip1.5mu \Varid{p}\mskip1.5mu]\mathbin{+\!\!\!+}\Varid{zs}} is stored in the array from index \ensuremath{\Varid{i}}, where \ensuremath{\Varid{ys'}} is a permutation of \ensuremath{\Varid{ys}}. The postcondition of \ensuremath{\Varid{ipartl}}, according to the specification \eqref{eq:ipartl-spec}, ends up with \ensuremath{\Varid{ys}} and \ensuremath{\Varid{zs}} stored consecutively. To connect the two conditions, we use a lemma that is dual to \eqref{eq:perm-write-swap}:
\begin{equation}
\begin{split}
&\ensuremath{\Varid{perm}\;\Varid{ys}\mathrel{\hstretch{0.7}{>\!\!>\!\!=}}\lambda \Varid{ys'}\to \Varid{writeList}\;\Varid{i}\;(\Varid{ys'}\mathbin{+\!\!\!+}[\mskip1.5mu \Varid{p}\mskip1.5mu])\mathrel{\supseteq}}\\
&\qquad\ensuremath{\Varid{writeList}\;\Varid{i}\;([\mskip1.5mu \Varid{p}\mskip1.5mu]\mathbin{+\!\!\!+}\Varid{ys})\mathbin{\hstretch{0.7}{>\!\!>}}\Varid{swap}\;\Varid{i}\;(\Varid{i}\mathbin{+}\#\!\!\;\Varid{ys})~~.}
\end{split} \label{eq:perm-write-swap-2}
\end{equation}
This is what the typical quicksort algorithm does: swapping the pivot with the last element of \ensuremath{\Varid{ys}}, and \eqref{eq:perm-write-swap-2} says that it is valid because that is one of the many permutations of \ensuremath{\Varid{ys}}. With~\eqref{eq:perm-write-swap-2} and \eqref{eq:ipartl-spec}, the specification can be refined to:
\begin{hscode}\SaveRestoreHook
\column{B}{@{}>{\hspre}l<{\hspost}@{}}%
\column{E}{@{}>{\hspre}l<{\hspost}@{}}%
\>[B]{}\Varid{writeList}\;\Varid{i}\;(\Varid{p}\mathbin{:}\Varid{xs})\mathbin{\hstretch{0.7}{>\!\!>}}{}\<[E]%
\\
\>[B]{}\Varid{ipartl}\;\Varid{p}\;(\Varid{i}\mathbin{+}\mathrm{1})\;(\mathrm{0},\mathrm{0},\#\!\!\;\Varid{xs})\mathrel{\hstretch{0.7}{>\!\!>\!\!=}}\lambda (\Varid{ny},\Varid{nz})\to \Varid{swap}\;\Varid{i}\;(\Varid{i}\mathbin{+}\Varid{ny})\mathbin{\hstretch{0.7}{>\!\!>}}{}\<[E]%
\\
\>[B]{}\Varid{iqsort}\;\Varid{i}\;(\#\!\!\;\Varid{ys})\mathbin{\hstretch{0.7}{>\!\!>}}\Varid{iqsort}\;(\Varid{i}\mathbin{+}\#\!\!\;\Varid{ys}\mathbin{+}\mathrm{1})\;(\#\!\!\;\Varid{zs})~~.{}\<[E]%
\ColumnHook
\end{hscode}\resethooks
%

In summary, we have derived:
\begin{hscode}\SaveRestoreHook
\column{B}{@{}>{\hspre}l<{\hspost}@{}}%
\column{13}{@{}>{\hspre}c<{\hspost}@{}}%
\column{13E}{@{}l@{}}%
\column{16}{@{}>{\hspre}l<{\hspost}@{}}%
\column{E}{@{}>{\hspre}l<{\hspost}@{}}%
\>[B]{}\Varid{iqsort}\mathbin{::}\Conid{MonadArr}\;\Conid{Elm}\;\Varid{m}\Rightarrow \Conid{Int}\to \Conid{Nat}\to \Varid{m}\;(){}\<[E]%
\\
\>[B]{}\Varid{iqsort}\;\Varid{i}\;\mathrm{0}{}\<[13]%
\>[13]{}\mathrel{=}{}\<[13E]%
\>[16]{}\{()\}{}\<[E]%
\\
\>[B]{}\Varid{iqsort}\;\Varid{i}\;\Varid{n}{}\<[13]%
\>[13]{}\mathrel{=}{}\<[13E]%
\>[16]{}\Varid{read}\;\Varid{i}\mathrel{\hstretch{0.7}{>\!\!>\!\!=}}\lambda \Varid{p}\to {}\<[E]%
\\
\>[16]{}\Varid{ipartl}\;\Varid{p}\;(\Varid{i}\mathbin{+}\mathrm{1})\;(\mathrm{0},\mathrm{0},\Varid{n}\mathbin{-}\mathrm{1})\mathrel{\hstretch{0.7}{>\!\!>\!\!=}}\lambda (\Varid{ny},\Varid{nz})\to {}\<[E]%
\\
\>[16]{}\Varid{swap}\;\Varid{i}\;(\Varid{i}\mathbin{+}\Varid{ny})\mathbin{\hstretch{0.7}{>\!\!>}}{}\<[E]%
\\
\>[16]{}\Varid{iqsort}\;\Varid{i}\;\Varid{ny}\mathbin{\hstretch{0.7}{>\!\!>}}\Varid{iqsort}\;(\Varid{i}\mathbin{+}\Varid{ny}\mathbin{+}\mathrm{1})\;\Varid{nz}~~.{}\<[E]%
\ColumnHook
\end{hscode}\resethooks

\section{Conclusions}
\label{sec:conclusions}

From a specification of sorting using the non-determinism monad, we have derived a pure quicksort for lists and a state-monadic quicksort for arrays.
We hope to demonstrate that the monadic style is a good choice as a calculus for program derivation that involves non-determinism.
One may perform the derivation in pointwise style, and deploy techniques that functional programmers have familiarised themselves with, such as pattern matching and induction on structures or on sizes.
When preferred, one can also work in point-free style with \ensuremath{(\mathrel{\hstretch{0.7}{>\!\!=\!\!\!>}})}.
Programs having other effects can be naturally incorporated into this framework.
The way we derive stateful programs echos how we, in Dijkstra's style, reason backwards from the postcondition.

A final note:
\ensuremath{(\mathrel{\hstretch{0.7}{>\!\!=\!\!\!>}})} and \ensuremath{(\dot{\subseteq})} naturally induce the notion of (left) factor, \ensuremath{(\mathbin{\backslash})\mathbin{::}(\Varid{a}\to \Varid{m}\;\Varid{b})\to (\Varid{a}\to \Varid{m}\;\Varid{c})\to \Varid{b}\to \Varid{m}\;\Varid{c}}, defined by the Galois connection:
\begin{align*}
  \ensuremath{\Varid{f}\mathrel{\hstretch{0.7}{>\!\!=\!\!\!>}}\Varid{g}\,\mathrel{\dot{\subseteq}}\,\Varid{h}~} &  \ensuremath{\mathrel{\equiv}~\Varid{g}\,\mathrel{\dot{\subseteq}}\,\Varid{f}\mathbin{\backslash}\Varid{h}~~.} \label{eq:left-factor-galois}
\end{align*}
Let \ensuremath{\Varid{h}\mathbin{::}\Varid{a}\to \Varid{m}\;\Varid{c}} be a monadic specification, and \ensuremath{\Varid{f}\mathbin{::}\Varid{a}\to \Varid{m}\;\Varid{b}} performs the computation half way,
then \ensuremath{\Varid{f}\mathbin{\backslash}\Varid{h}} is the most non-deterministic (least constrained) monadic program that, when ran after the postcondition set up by \ensuremath{\Varid{f}}, still meets the result specified by \ensuremath{\Varid{h}}. With \ensuremath{(\mathbin{\backslash})}, \ensuremath{\Varid{ipartl}} and \ensuremath{\Varid{iqsort}} can be specified by:
\begin{hscode}\SaveRestoreHook
\column{B}{@{}>{\hspre}l<{\hspost}@{}}%
\column{4}{@{}>{\hspre}l<{\hspost}@{}}%
\column{16}{@{}>{\hspre}l<{\hspost}@{}}%
\column{E}{@{}>{\hspre}l<{\hspost}@{}}%
\>[4]{}\Varid{ipartl}\;\Varid{p}\;\Varid{i}{}\<[16]%
\>[16]{}\mathrel{\dot{\subseteq}}\Varid{write3L}\;\Varid{i}\mathbin{\backslash}((\Varid{second}\;\Varid{perm}\mathbin{\cdot}\Varid{partl}\;\Varid{p})\mathrel{\hstretch{0.7}{>\!\!=\!\!\!>}}\Varid{write2L}\;\Varid{i})~~,{}\<[E]%
\\
\>[4]{}\Varid{iqsort}\;\Varid{i}{}\<[16]%
\>[16]{}\mathrel{\dot{\subseteq}}\Varid{writeL}\;\Varid{i}\mathbin{\backslash}(\Varid{slowsort}\mathrel{\hstretch{0.7}{>\!\!=\!\!\!>}}\Varid{writeList}\;\Varid{i})~~.{}\<[E]%
\ColumnHook
\end{hscode}\resethooks
In relational calculus, the {\em factor} is an important operator that is often associated with weakest precondition.
We unfortunately cannot cover it due to space constraints.

\paragraph{Acknowledgements}
The authors would like to thank Jeremy Gibbons for the valuable discussions during development of this work. 

\bibliographystyle{splncs04}

\end{document}